\documentclass[11pt,a4paper]{article}
\usepackage{jheppub}

\usepackage{multirow, graphicx,amssymb,url,mathrsfs,amsmath}
\usepackage{wrapfig,boxedminipage,setspace,subfigure,epsfig}
\usepackage{amsxtra,amstext,latexsym,dsfont,amsfonts}
\usepackage{color,eucal}
\usepackage[dvipsnames]{xcolor}
\usepackage{float}
\usepackage{slashed,comment}
\usepackage{kotex}

\usepackage{tikz}
\usetikzlibrary{calc,patterns,angles,quotes}
\usetikzlibrary{decorations.pathreplacing,decorations.markings,snakes}

\usepackage{tabularx, array}
\usepackage{contour}
\newcolumntype{L}[1]{>{\raggedright\arraybackslash}p{#1}}
\newcolumntype{C}[1]{>{\centering\arraybackslash}p{#1}}
\newcolumntype{R}[1]{>{\raggedleft\arraybackslash}p{#1}}

\renewcommand{\l}{\lambda}

\newcommand{\nn}{\nonumber}

\usepackage{xparse}
\ExplSyntaxOn
\NewDocumentCommand{\HS}{m}
 {
  \seq_set_split:Nnn \l_tmpa_seq { ~ } { #1 }
  \seq_map_inline:Nn \l_tmpa_seq { \contour{green}{##1} ~ } \unskip
 }
\ExplSyntaxOff

\title{Renyi reflected entropy and entanglement wedge cross section with cosmic branes in AdS/BCFT}

\author[a]{Byoungjoon Ahn,}
\author[b]{Sang-Eon Bak,}
\author[a,c]{Keun-Young Kim,}
\author[d,e]{and Mitsuhiro Nishida}

\emailAdd{bjahn123@gist.ac.kr}
\emailAdd{sbak2@asu.edu}
\emailAdd{fortoe@gist.ac.kr}
\emailAdd{mnishida124@gmail.com}

\affiliation[a]{Department of Physics and Photon Science, Gwangju Institute of Science and Technology, \\
123 Cheomdan-gwagiro, Gwangju 61005, Korea}
\affiliation[b]{Department of Physics, Arizona State University, Tempe, AZ 85281, USA}
\affiliation[c]{Research Center for Photon Science Technology, Gwangju Institute of Science and Technology, \\
123 Cheomdan-gwagiro, Gwangju 61005, Korea}
\affiliation[d]{Department of Physics, Pohang University of Science and Technology, Pohang 37673, Korea}
\affiliation[e]{National Institute of Technology, Yuge College, Ehime 794-2593, Japan}

\abstract{
In this study, we calculate the $m-1$ correction to the reflected entropy for two adjacent intervals on a half-infinite line within the AdS$_3$/BCFT$_2$ framework, where $m$ is a Renyi index for a canonical purification. We utilize the doubling trick and compute the leading terms in the large central charge expansion of correlation functions in the holographic BCFT. In the corresponding AdS space with an end of the world brane, we analyze the entanglement wedge cross section, the dual counterpart of reflected entropy. This AdS/BCFT setup allows us to explore a richer set of phases in the entanglement wedge cross section. The $m-1$ correction in the holographic BCFT manifests as modifications in the entanglement wedge cross section induced by cosmic branes. For the adjacent intervals anchored to the boundary of BCFT, we show the duality between the entanglement wedge cross section with the backreaction from a cosmic brane and Renyi reflected entropy at all orders in $m-1$. Furthermore, by analyzing the entanglement wedge cross section for general adjacent intervals, we provide guidance for an $\epsilon$-expansion of five-point functions in the holographic CFT, where $\epsilon$ is the rescaled conformal dimension by the central charge.
}

\begin{document}
\maketitle

%
\section{Introduction}

AdS/CFT correspondence \cite{Maldacena:1997re, Gubser:1998bc, Witten:1998qj} has become a powerful and intuitive framework for examining physical systems, leading to numerous studies since its proposal. 
Notably, it has been applied in various research areas such as quantum chromodynamics, condensed matter theory, and quantum information. The Ryu-Takayanagi formula \cite{Ryu:2006bv, Ryu:2006ef} was also a particularly significant breakthrough, making these studies more accessible. This formula allowed entanglement entropy calculated in field theory to be expressed as geometric quantities in the corresponding gravitational system. Through its various applications, AdS/CFT not only uncovers new physics but also serves as a valuable tool for expanding calculations due to its inherent duality.

There are various physical quantities used to measure the entanglement of quantum systems.
Among them, entanglement entropy is a well-defined entanglement measure for pure states. However, it is not suitable as a measure for the entanglement of mixed states \cite{Horodecki:2009zz, Carisch:2022ufd}. The entanglement structure of mixed states is more complex, and there may be different ways to decompose correlations in a given mixed state density matrix. It is challenging to distinguish classical correlations from quantum entanglement in statistical mixtures. Several quantities have been proposed to study entanglement in mixed states, such as entanglement of purification \cite{Takayanagi:2017knl, Nguyen:2017yqw}, reflected entropy \cite{Dutta:2019gen}, odd entanglement entropy \cite{Tamaoka:2018ned}, and entanglement negativity \cite{Vidal:2002zz, Plenio:2005cwa, Calabrese:2012ew, Kudler-Flam:2018qjo}.

Our primary interest lies in reflected entropy \cite{Dutta:2019gen,Jeong:2019xdr,Akers:2021pvd,Akers:2022max}, a quantity designed to probe entanglement in mixed states.
Although it is not a correlation measure for a generic state \cite{Hayden:2023yij}, reflected entropy provides valuable insight into mixed state entanglement, particularly in the context of tripartite systems ~\cite{Hayden:2021gno, Akers:2019gcv}. In this work, we investigate reflected entropy within the AdS$_3$/CFT$_2$ framework at large central charge $c$ \cite{Hartman:2013mia,Fitzpatrick:2014vua,Perlmutter:2015iya,Anous:2016kss},
where its monotonicity has been established for certain classes of states in holographic field theories~\cite{Dutta:2019gen}.
Through this approach, we aim to extend the scope of calculations and strengthen the validity of AdS/CFT correspondence. We revisit and integrate previous studies, expanding calculations as far as possible. Specifically, the reason why we focus on reflected entropy is that it exhibits a different structure compared to entanglement entropy. Both entanglement entropy and reflected entropy can be expressed in terms of correlation functions in conformal field theory (CFT) \cite{Calabrese:2004eu,Calabrese:2009qy}. However, from the perspectives of $1/c$ expansion and conformal blocks, reflected entropy offers a wider range of scenarios.

Moreover, by considering the effects of boundaries in CFT \cite{Cardy:2004hm,Takayanagi:2011zk,Fujita:2011fp,Nozaki:2012qd,Jensen:2013lxa}, we can explore a diverse array of possible calculations. 
Reflected entropy \cite{Dutta:2019gen} in holographic CFT has been studied using the replica trick, and it is represented in terms of the four-point correlation functions of the twist operators for two disjoint intervals. Depending on the cross-ratio, one obtains either a zero or non-zero value of the reflected entropy, so there is only one non-trivial phase of reflected entropy for two disjoint intervals. However, if we introduce the boundary of the holographic CFT, the existence of the boundary compels us to use the doubling trick \cite{Cardy:1984bb}, which provides a method to evaluate $N$-point correlation functions of boundary conformal field theory (BCFT) on a half-plane by using $2N$-point correlation functions of CFT on the full plane \cite{Sully:2020pza}. By employing the doubling trick, one ends up with a greater number of non-trivial phases of reflected entropy \cite{Li:2021dmf,BasakKumar:2022stg}.

The gravity dual of reflected entropy is known as the entanglement wedge cross section in AdS spacetime \cite{Takayanagi:2017knl,Nguyen:2017yqw}. When we consider the entanglement wedge \cite{Czech:2012bh, Wall:2012uf, Headrick:2014cta} associated with two subregions on the CFT, only one non-trivial phase of the entanglement wedge cross section emerges. However, in the context of BCFT, the dual geometry incorporates the end of the world brane \cite{Karch:2000gx, Takayanagi:2011zk, Fujita:2011fp, Nozaki:2012qd, Geng:2021iyq,Geng:2022dua,Geng:2024xpj}.  The presence of the end of the world brane allows the entanglement wedge associated with two subregions to involve a larger set of phases of the entanglement wedge cross section. For recent developments associated with reflected entropy and entanglement wedge cross section in AdS/BCFT and similar settings, see \cite{Sahraei:2021wqn, Chen:2022fte, Kusuki:2022bic, Vasli:2022kfu, Afrasiar:2022fid,Afrasiar:2023jrj,Basu:2023wmv,Basu:2023jtf, Lin:2023ajt,Basak:2023bnc}.

In this work, we generalize the analysis of the entanglement wedge cross section with corrections for the replica index in AdS/CFT \cite{Jeong:2019xdr} to that in AdS/BCFT, which has a richer set of phases.
The reflected entropy is computed using the replica trick by taking the limits of the replica indices $n \rightarrow 1$ and $m \rightarrow 1$. One can relax this limit by making $m$ slightly different from 1. 
The $m-1$ correction to the reflected entropy on the boundary sides is computed through the perturbative expansions of the conformal block in the semi-classical limit \cite{Fitzpatrick:2014vua,Perlmutter:2015iya,Anous:2016kss}.
On the other hand, the corresponding correction to the entanglement wedge cross section manifests through the backreaction of cosmic branes to the geometry, requiring us to calculate the geodesics in the geometry involving the conical singularity \cite{Hung:2011nu,Dong:2016fnf}.
In this sense, without assuming the GKP-Witten relation, the bulk and boundary calculations are quite different and the alignment of these two independent calculations is not straightforward.

Our goal is to find the duality in the $m-1$ correction on the boundary side and the presence of the backreaction on the bulk side. This will allow us to further solidify the correspondence between the bulk and boundary descriptions in the AdS/BCFT framework and explore the richer set of phases that arise in this setup.

This paper is organized as follows. In Sec.~\ref{sec:ReflectedEntropy}, we begin with the BCFT calculation. In Sec.~\ref{conceptofRE}, we introduce the assumptions of our calculations and review the reflected entropy in BCFT employing the doubling trick. We then calculate the $m-1$ correction to the reflected entropy in Sec.~\ref{sec_2.2}. Moving into the gravity calculation in Sec.~\ref{sec_3}, we review the entanglement wedge cross section without backreaction in Sec.~\ref{sec_3.1.2}. By using the conformal transformation, we calculate the entanglement wedge cross section involving backreaction from the cosmic brane in Sec.~\ref{sec_3.2}. To extend the scope of the calculation, we calculate the entanglement wedge cross section in a more general configuration of boundary intervals in Sec.~\ref{sec 4.1} and show the consistency with the $m-1$ corrections in the AdS/CFT setup in Sec.~\ref{sec_4.2}. We conclude this paper by summarizing our results and outlook in Sec.~\ref{sec_conclusion}.

%
\section{Reflected entropy with $m-1$ corrections}\label{sec:ReflectedEntropy}
In this section, we aim to calculate the $m-1$ corrections to the reflected entropy in AdS$_3$/BCFT$_2$ for various configurations. This analysis helps establish a duality between the $m-1$ corrections to the reflected entropy in BCFT$_2$ and the backreaction effects on the entanglement wedge cross section in AdS$_3$, specifically considering the end of the world brane. We begin by reviewing the concept of reflected entropy in quantum theories with finite-dimensional Hilbert spaces. Next, we introduce reflected entropy within the framework of CFT$_2$, employing the replica trick. The reflected entropy is then expressed through correlation functions of twist operators, allowing for $m-1$ corrections via the large central charge expansion. In BCFT$_2$, which include a boundary at the origin, a broader set of correlation functions is necessary to compute the reflected entropy for each subregion configuration. The richer structure of BCFT$_2$ becomes evident through the doubling trick, which effectively maps the correlation functions in BCFT$_2$ to those in CFT$_2$. Employing these techniques in the the boundary context, we derive the reflected entropy with $m-1$ corrections.

\subsection{Reflected entropy in BCFT$_2$}\label{conceptofRE}

\paragraph{Reflected entropy.}

In our work, we focus on reflected entropy for entanglement in mixed states. Reflected entropy is defined for a bipartite quantum system $A \cup B$ and a mixed state $\rho_{AB}$. The authors of \cite{Dutta:2019gen} introduced a canonical purification to define a purified state $\left|\sqrt{\rho_{A B}}\right\rangle$, which does not require us to choose any specific basis in the Hilbert space.
The mixed state $\rho_{AB}$ is described as a reduced density matrix of the purified state $\left|\sqrt{\rho_{A B}}\right\rangle$ in a doubled Hilbert space:
\begin{align}
\begin{split}
\rho_{A B} = \operatorname{Tr}_{\mathcal{H}_A^{\star} \otimes \mathcal{H}_B^{\star}}\left|\sqrt{\rho_{A B}}\right\rangle\left\langle\sqrt{\rho_{A B}}\right| \,,
\end{split}
\end{align}
where 
\begin{align}
\begin{split}
\left|\sqrt{\rho_{A B}}\right\rangle \in  \left(\mathcal{H}_A \otimes \mathcal{H}_A^{\star}\right) \otimes\left(\mathcal{H}_B \otimes \mathcal{H}_B^{\star}\right) =: \mathcal{H}_{A A^{\star} B B^{\star}} \,.
\end{split}
\end{align}
Then, reflected entropy is defined as the von Neumann entropy of another mixed state $\rho_{AA^\star}$ as
\begin{align}\label{RE}
\begin{split}
S_R(A: B) := \,&-\text{Tr}_{\mathcal{H}_A\otimes\mathcal{H}^\star_A}\left[\rho_{AA^\star}\log\rho_{AA^\star}\right] \,,\\
\rho_{AA^\star} := \,&\text{Tr}_{\mathcal{H}_B\otimes\mathcal{H}^\star_B}|\sqrt{\rho_{AB}}\rangle\langle\sqrt{\rho_{AB}}| \,,
\end{split}
\end{align}
which can be interpreted as the entanglement entropy of the reduced density matrix $\rho_{AA^\star}$. Reflected entropy is a useful quantity for analyzing entanglement in mixed states, which reflects both quantum and classical correlations between $A$ and $B$, though the reflected entropy of some states violates a condition for a measure of correlations.

\paragraph{Replica trick for the reflected entropy.}
The replica trick serves as an effective method for computing entanglement entropy within the path integral framework, as outlined in \cite{Calabrese:2004eu}. 
The entanglement entropy is calculated from the limit of Renyi entropy generalized with respect to replica index $n$. The $n$th power of the density matrix is obtained by a path integral over the replica manifold. Although the Renyi entropy is traditionally defined for integer values of $n$, analytic continuation is assumed when taking the $n \rightarrow 1$ limit.
This approach is also applicable in calculating reflected entropy. 

The replica partition function can be characterized in two distinct ways. The first involves a path integral over a replica manifold with branch cuts across intervals. The second approach expresses it as correlation functions of twist operators, which are defined in an $n$-fold product theory on the original manifold. These fields are positioned at the endpoints of these intervals and meet a specific boundary condition: the rotation around the twist operator must align with the rotation around interval's endpoints on the replica manifold.

The state $\left|\sqrt{\rho_{A B}}\right\rangle$ is generalized for two replica indices, $n$ and $m$ (See \cite{Dutta:2019gen, Jeong:2019xdr} for details) as follows.\footnote{Here, $n\in \mathbb{Z}^{+}$ and $m\in 2 \mathbb{Z}^{+}$.}
First, a canonical purification of $\rho_{A B}^m$, $| \psi_m \rangle$, is written as
\begin{equation}
| \psi_m \rangle :=\frac{1}{\sqrt{\operatorname{Tr} \rho_{A B}^m}}	 |\rho_{A B}^{m / 2} \rangle,
\end{equation}
with the normalization:
\begin{equation}
\operatorname{Tr}_{\mathcal{H}_A^{\star} \otimes \mathcal{H}_B^{\star}}\left|\psi_m\right\rangle\left\langle\psi_m\right|=\frac{\rho_{A B}^m}{\operatorname{Tr} \rho_{A B}^m}	.
\end{equation}
Then, the extended reflected entropy with respect to the replica indices is defined as follows:
\begin{align}\label{Renyi reflected entropy definition}
S_n\left(A A^{\star}\right)_{\psi_m} & :=\frac{1}{1-n} \log \operatorname{Tr}_{\mathcal{H}_A \otimes \mathcal{H}_A^{\star}}\left(\rho_{A A^{\star}}^{(m)}\right)^n, \\
\rho_{A A^{\star}}^{(m)} & :=\operatorname{Tr}_{\mathcal{H}_B \otimes \mathcal{H}_B^{\star}}\left|\psi_m\right\rangle\left\langle\psi_m\right|,
\end{align}
where $\rho_{A A^{\star}}^{(m)}$ is the reduced density matrix by tracing over the subsystems $B$ and $B^{\star}$.
After the analytic continuation of the replica indices $n, m$, the extended reflected entropy converges to reflected entropy when $n \rightarrow 1$ and $m \rightarrow 1$:
\begin{equation}
\lim _{n, m \rightarrow 1} S_n\left(A A^{\star}\right)_{\psi_m}=S_R(A: B) .	
\end{equation}
In this paper, we focus on the $m-1$ correction of $S_R^{(m)}(A:B)$ defined by
\begin{equation}\label{DefRenyiRE}
S_R^{(m)}(A:B):=\lim _{n\rightarrow 1} S_n\left(A A^{\star}\right)_{\psi_m},
\end{equation}
and the Renyi reflected entropy in this paper means $S_R^{(m)}(A:B)$.\footnote{The Renyi reflected entropy in this paper is equivalent to the $(m,1)-$Renyi reflected entropy in \cite{Akers:2022max}.} The Renyi reflected entropy can be represented in terms of the partition functions 
\begin{align}\label{eq_SR with partition}
     S_R^{(m)}(A:B)\equiv \lim _{n \rightarrow 1} \frac{1}{1-n} \log{\frac{Z_{mn}}{(Z_m)^n}}\,,
\end{align}
where we defined the un-normalized partition function:
\begin{align}\label{eq_def partition function}
    Z_{mn}:=\operatorname{Tr}_{\mathcal{H}_A \otimes \mathcal{H}_A^{\star}}\left(\operatorname{Tr}_{\mathcal{H}_B \otimes \mathcal{H}_B^{\star}}\left|\rho_{A B}^{m / 2}\right\rangle\left\langle\rho_{A B}^{m / 2}\right|\right)^n,\quad Z_m:=Z_{m n}|_{n \rightarrow 1}\,.
\end{align}

One can think of the extended Renyi entropy \eqref{Renyi reflected entropy definition} as the correlation function of $n$-fold twist operators with a boundary condition for the Euclidean branched manifold. For example, if we consider two intervals $A=\left[x_1, x_2\right]$ and $B=\left[x_3, x_4\right]$ with $x_1<x_2<x_3<x_4$, the extended reflected entropy is computed via \cite{Dutta:2019gen, Jeong:2019xdr}
\begin{equation}
S_n\left(A A^{\star}\right)_{\psi_m} = \frac{1}{1-n} \log  \frac{\langle\sigma_{g_A}(x_1) \sigma_{g_A^{-1}}(x_2) \sigma_{g_B}(x_3) \sigma_{g_B^{-1}}(x_4)\rangle_{\mathrm{CFT}^{\otimes mn}}}{\left(\langle\sigma_{g_m}(x_1) \sigma_{g_m^{-1}}(x_2) \sigma_{g_m}(x_3) \sigma_{g_m^{-1}}(x_4)\rangle_{\mathrm{CFT}^{\otimes m}}\right)^n} \,,
\end{equation}
where $\mathrm{CFT}^{\otimes m n}$ represents the product theory on two-dimensional flat spacetime, which contains $mn$ replica fields for the $mn$ replica sheets.
Here, $\sigma_{g_m}$ is the usual twist operator for $m$ replica sheets \cite{Calabrese:2009qy, Calabrese:2004eu} with the conformal dimension $h_{g_m}=\frac{c}{24}\left(m-\frac{1}{m}\right)$, and $\sigma_{g_A}, \sigma_{g_A^{-1}}, \sigma_{g_B}$, and $\sigma_{g_B^{-1}}$ are the twist operators located at the end of each interval obeying a specific boundary condition (See \cite{Dutta:2019gen, Jeong:2019xdr} for details). Their conformal dimensions are determined as
\begin{align}\label{conformaldimension}
    h_{g_A}=h_{g_B}=\frac{n c}{24}\left(m-\frac{1}{m}\right), \qquad h_{g_B g_A^{-1}}=\frac{2 c}{24}\left(n-\frac{1}{n}\right)\,,
\end{align}
where $h_{g_A^{-1}}=h_{g_A}$ and $h_{g_B^{-1}}=h_{g_B}$, and $h_{g_B g_A^{-1}}$ is the conformal dimension of a twist operator $\sigma_{g_B g_A^{-1}}$ in the limit $x_2\to x_3$.

\paragraph{Large central charge expansion.}
In CFT$_2$ with Virasoro symmetry, a four-point function of primary operators can be expanded using Virasoro conformal blocks.
Assuming a large central charge and considering specific conditions of the CFT spectrum\footnote{Such a special class of theory that obeys those conditions is dubbed as holographic CFT in this paper, following \cite{Hartman:2013mia, Dutta:2019gen}.}, both the Renyi entropy and the entanglement entropy can be computed without requiring the detailed specifics of the CFT.
In this regime, the dominant contribution to the Renyi entropy in the large central charge limit is captured by the Virasoro conformal block for the vacuum state.
This contribution can be determined by applying a trivial monodromy condition to a relevant differential equation. 
We can generalize the Virasoro conformal block, which relies solely on the algebra, to cases involving more than four operators.
In this paper, we evaluate the leading term of the perturbative expansion in $1/c$ and do not take into account non-perturbative corrections \cite{Akers:2021pvd, Akers:2022max}.
For further details, refer to \cite{Hartman:2013mia, Anous:2016kss}.

\paragraph{Doubling trick in BCFT$_2$.}
The doubling trick, introduced by Cardy \cite{Cardy:1984bb}, is a technique that employs mirror images in a BCFT$_2$ to transform correlation functions of local operators in a BCFT$_2$ into correlation functions that involve both the local operators and their mirror images in a CFT$_2$. This approach is valid because the correlation functions in the BCFT$_2$ satisfy a linear combination of the conformal Ward identities, for transformations that map the boundary to itself, of the CFT$_2$.

Since the Virasoro symmetry algebra of a BCFT is identical to that of a chiral CFT on the entire plane, the kinematics of a BCFT defined on the upper-half plane (UHP) are directly related to those of a chiral CFT on the full plane \cite{Sully:2020pza}.
On the UHP, the correlators of bulk operators $\mathcal{O}_{h \bar{h}}$ with conformal weights $(h, \bar{h})$ must have the same functional form as the correlators of operators $\mathcal{O}_{h}$ and $\mathcal{O}_{\bar{h}}$ in a chiral CFT on the whole plane.
\begin{equation}
    \left\langle \mathcal{O}_{h_1, \bar{h}_1}(z_1, \bar{z}_1) ... \mathcal{O}_{h_n, \bar{h}_n}(z_n, \bar{z}_n) \right\rangle_{\mathrm{UHP}} \sim  \left\langle \mathcal{O}_{h_1}(z_1) ... \mathcal{O}_{h_n}(z_n)  \mathcal{O}_{\bar{h}_1}( \bar{z}_1)... \mathcal{O}_{\bar{h}_n}( \bar{z}_n) \right\rangle
\end{equation}
This correspondence allows for a comparison between the kinematics of $n$-point correlation functions in a BCFT on the half-plane and $2n$-point functions in a chiral CFT on the full plane.

Through the doubling trick, a scalar primary operator with weights $(h, \bar{h})$ in a BCFT can have a non-vanishing one-point function.
In this case, the one-point function $\left\langle\mathcal{O}_{h, \bar{h}}(z, \bar{z})\right\rangle_{\mathrm{UHP}}$ can be written as a chiral two-point function $\left\langle\mathcal{O}_h(z) {\mathcal{O}}_{\bar{h}}\left(z^*\right)\right\rangle$, 
leading to the equation
\begin{equation}\label{tpBCFT}
\left\langle\mathcal{O}_{h, \bar{h}}(z, \bar{z})\right\rangle_{\mathrm{UHP}}=\frac{\mathcal{C}_{\mathcal{O}}}{\left|z-z^*\right|^{2 h}}=\frac{\mathcal{C}_{\mathcal{O}}}{|2 x|^{2h}}.
\end{equation}
where $z=t+ix$.
Here, the coefficient $\mathcal{C}_{\mathcal{O}}$ in the one-point function becomes a physical parameter that typically depends on both the operator and the boundary condition such as the existence of the boundary degrees of freedom.

\paragraph{Reflected entropy in BCFT.} Consider the two intervals defined in the 2-dimensional CFT with a boundary at the origin as
\begin{align}
    A=\left[0, x_1\right] \,, \quad B=\left[x_1, x_2\right] \,.
\end{align}
By using the replica trick, the reflected entropy is computed based on the correlation functions of the twist operators
\begin{align}\label{eq_simple adj ent}
    S_R(A: B) = \lim_{m, n \rightarrow 1} \frac{1}{1-n} \log  \frac{\left\langle \sigma_{g_B g_A^{-1} }(x_1) \sigma_{g_B^{-1}}(x_2) \right\rangle_{\mathrm{BCFT}^{\otimes mn}}}{\left(\left\langle \sigma_{g_m}(x_2) \right\rangle_{\mathrm{BCFT}^{\otimes m}}\right)^n}\,, 
\end{align}
where the denominator became a one-point function since
\begin{align}\label{eq_simple adj ent denom}
\lim _{n \rightarrow 1}\left\langle \sigma_{g_B g_A^{-1} }(x_1) \sigma_{g_B^{-1}}(x_2) \right\rangle_{\mathrm{BCFT}^{\otimes mn}}=\big\langle \sigma_{g_m}(x_2) \big\rangle_{\mathrm{BCFT}^{\otimes m}}\,,
\end{align}
for $\lim _{n \rightarrow 1} h_{g_B g_A^{-1}}=0$ and $\lim _{n \rightarrow 1} h_{g_B}=h_{g_m}$. The conformal dimensions for each twist operator are given by (\ref{conformaldimension}) and $h_{g_m}=\frac{c}{24}\left(m-\frac{1}{m}\right)$.

By applying the doubling trick, the reflected entropy, as outlined in (\ref{eq_simple adj ent}), becomes
\begin{align}\label{eq_simple_ent_in_doubling}
    S_R(A:B) = \lim_{m,n \to 1} \frac{1}{1-n} \log \frac{\left\langle \sigma_{g_B}(-x_2) \sigma_{g_A g_B^{-1} }(-x_1) \sigma_{g_B g_A^{-1} }(x_1) \sigma_{g_B^{-1}}(x_2) \right\rangle_{\mathrm{CFT}^{\otimes mn}}}{\left(\left\langle \sigma_{g_m}(-x_2) \sigma_{g_m^{-1}}(x_2) \right\rangle_{\mathrm{CFT}^{\otimes m}}\right)^n}\,.
\end{align}
Thus, we can calculate the reflected entropy in BCFT using the procedures established for CFT. It is important to note that, due to the doubling trick, we must compute correlation functions in chiral CFT. A straightforward method for chiral CFT in this context involves calculating the holomorphic and antiholomorphic parts and then halving the final result. In the configuration of two adjacent intervals, the system may exhibit two possible phases depending on the relative sizes of the intervals $A$ and $B$.

\subparagraph{(i) Phase I.} 

\begin{figure}[]
    \centering
\includegraphics[width=0.71\textwidth]{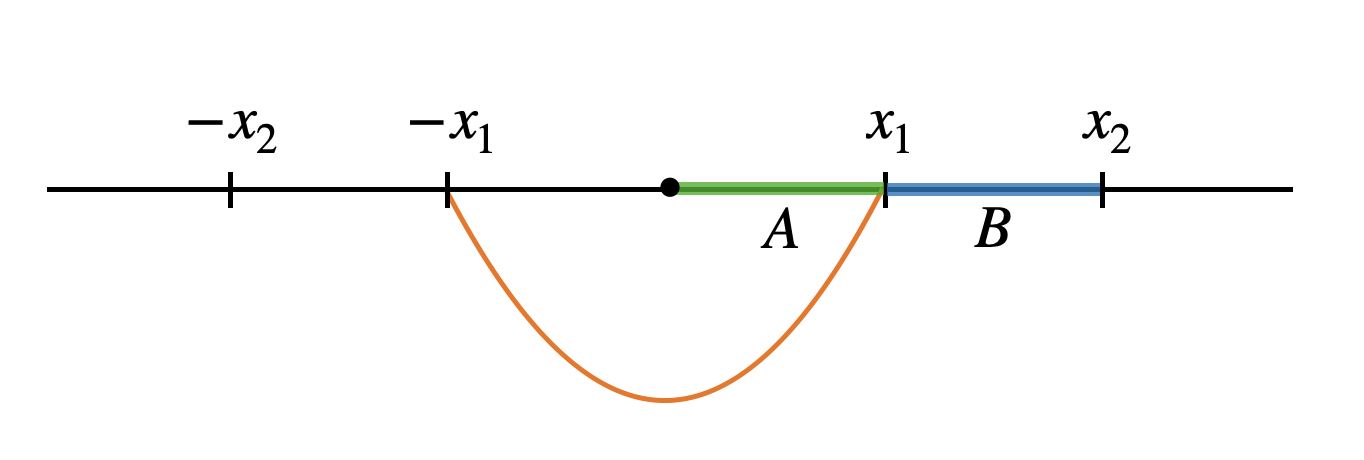}
    \caption{The orange line represents the OPE channel in phase I for the configuration of two adjacent intervals $A=[0,x_1], B=[x_1,x_2]$ where two copies of BCFT become a chiral CFT by using the doubling trick.}
    \label{fig1}
\end{figure}

When the size of the interval $A$ is much smaller than that of the interval $B$, the four-point vacuum conformal block in the OPE channel for two pairs, $\sigma_{g_A g_B^{-1} }(-x_1) \sigma_{g_B g_A^{-1} }(x_1)$ and $\sigma_{g_B}(-x_2) \sigma_{g_B^{-1}}(x_2)$, may be dominant for the numerator in equation (\ref{eq_simple_ent_in_doubling}). In addition, if the OPE for one of the pairs is simplified by an assumption that the twist operator can be regarded as the identity operator at $m, n \rightarrow 1$, the four-point function may be represented by lower-point functions. Under the above assumptions, we treat $\sigma_{g_B}(-x_2)$ and $\sigma_{g_B^{-1}}(x_2)$ as the identity operators at $m\to1$ and decouple it from $\sigma_{g_A g_B^{-1} }(-x_1) \sigma_{g_B g_A^{-1} }(x_1)$.\footnote{If we treat both $\sigma_{g_A g_B^{-1} }(-x_1)$ and $\sigma_{g_B g_A^{-1}}(x_1)$ as the identity operators at $n\to1$, we cannot obtain a nonzero result. Strictly speaking, one must pay attention to the order of the limits $m\to1$ and $n\to1$ \cite{Kusuki:2019evw}, but we are not concerned with that in this calculation. At least, our CFT calculation is consistent with the gravity computation in the next section. Confirming the assumption of the dominant channel is a future work.} Then, the numerator in equation (\ref{eq_simple adj ent}) is factorized by
\begin{align}
    \left\langle \sigma_{g_B g_A^{-1} }(x_1) \sigma_{g_B^{-1}}(x_2) \right\rangle_{\mathrm{BCFT}^{\otimes mn}} = \left\langle \sigma_{g_B g_A^{-1}}(x_1) \right\rangle_{\mathrm{BCFT}^{\otimes mn}} \left\langle \sigma_{g_B^{-1}}(x_2) \right\rangle_{\mathrm{BCFT}^{\otimes mn}}\,.
\end{align}
Since the one-point function at $x_2$ is canceled by the denominator after taking the limit as $n$ approaches 1, the reflected entropy is determined by the one-point function at $x_1$. By using the doubling trick, the one-point function at $x_1$ is computed by the two-point function as shown in Fig.~\ref{fig1}.
\begin{align}\label{eq_SR for zero phase 1}
\begin{split}
    S_R(A: B) &= \lim_{m, n \rightarrow 1} \frac{1}{1-n} \log \left\langle \sigma_{g_A g_B^{-1} }(-x_1) \sigma_{g_B g_A^{-1} }(x_1) \right\rangle_{\mathrm{CFT}^{\otimes mn}} \\
    &= \lim_{m, n \rightarrow 1} \frac{1}{(1-n)} \log  \frac{e^{(1-n) S_{\mathrm{bdy}}} \delta_x^{2h_{g_B g_A^{-1}}}}{(2x_1)^{2h_{g_B g_A^{-1} }}}\,,
\end{split}
\end{align}
where we only took the holomorphic part, accounting for the chiral CFT.
If we have the boundary degrees of freedom, the coefficient $\mathcal{C}_{\mathcal{O}}$ in \eqref{tpBCFT} involves the boundary entropy $S_{\mathrm{bdy}}$ (or g-function) \cite{Affleck:1991tk, Cardy:2004hm, Takayanagi:2011zk, Fujita:2011fp}.
UV-cutoff, denoted as $\delta_x$, is introduced to make the final result dimensionless \cite{Calabrese:2004eu}.
We will introduce $\delta_x$ in subsequent calculations as appropriate.
Consequently, the reflected entropy for phase I is calculated as
\begin{align}\label{eq_SR phase 1}
    S_R(A: B)=\frac{c}{3} \log \left(\frac{2 x_1}{\delta_x}\right) \,,
\end{align}
where we set $S_{\mathrm{bdy}}=0$. We also note that throughout this paper, we assume the boundary entropy, $S_{\mathrm{bdy}}$, is zero.

\subparagraph{(ii) Phase II.} 
\begin{figure}[]
    \centering
    \includegraphics[width=1\textwidth]{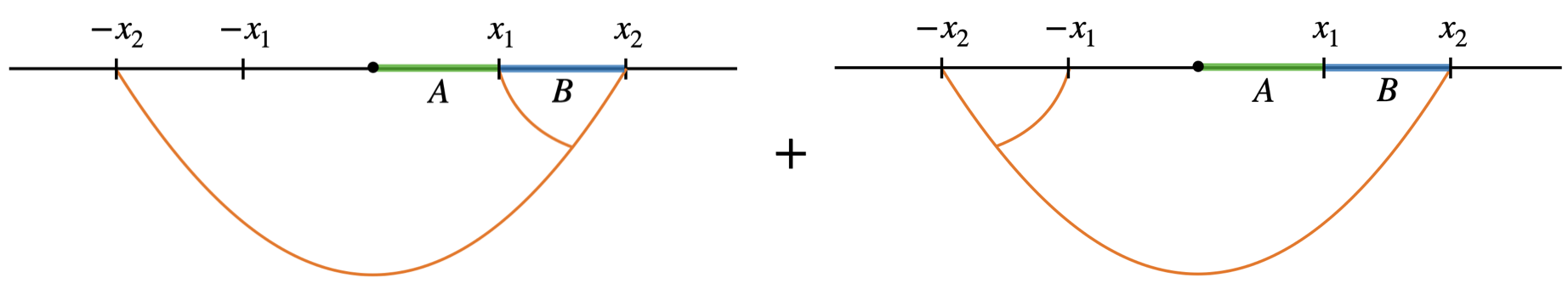}
    \caption{Schematic figure representing the OPE channel for phase II in the case of adjacent intervals that are contacted with the boundary of BCFT. In this case, the two configurations of the three-point function in CFT contribute in principle to the two-point function in BCFT when we take the doubling trick.}
    \label{fig2}
\end{figure}

When the size of the interval $A$ is much larger than that of the interval $B$, the four-point conformal block in the OPE channel for two pairs, $\sigma_{g_A g_B^{-1} }(-x_1) \sigma_{g_B}(-x_2)$ and $ \sigma_{g_B g_A^{-1} }(x_1)\sigma_{g_B^{-1}}(x_2)$, may be dominant for the numerator in equation (\ref{eq_simple_ent_in_doubling}). In phase II, we assume that either $\sigma_{g_A g_B^{-1} }(-x_1)$ or $\sigma_{g_B g_A^{-1} }(x_1)$ can be treated as the identity operator at $n\to1$. For example, if we treat $\sigma_{g_A g_B^{-1} }(-x_1)$ as the identity operator, the numerator in equation (\ref{eq_simple adj ent}) can be evaluated by
\begin{align}\label{phaseII}
    \left\langle \sigma_{g_B g_A^{-1}}(x_1) \sigma_{g_B^{-1}}(x_2) \right\rangle_{\mathrm{BCFT}^{\otimes mn}} = \left\langle \sigma_{g_B}(-x_2) \sigma_{g_B g_A^{-1} }(x_1) \sigma_{g_B^{-1}}(x_2) \right\rangle_{\mathrm{CFT}^{\otimes mn}}\,,
\end{align}
where we used the doubling trick. 
To be precise, we should consider the contributions from equation (\ref{phaseII}) and another three-point function $\langle\sigma_{g_B}(-x_2)\sigma_{g_A g_B^{-1} }(-x_1)\sigma_{g_{B}^{-1}}(x_2)\rangle$ in the chiral CFT (See Fig.~\ref{fig2}). Instead of analyzing them, we evaluate both holomorphic and antiholomorphic parts of equation (\ref{phaseII}) for the full correlator, not the chiral part. Thanks to the symmetric configurations, our computation of the full correlator (\ref{phaseII}) matches with the chiral parts of the two configurations in Fig.~\ref{fig2}.
Consequently, the reflected entropy is determined as follows
\begin{align}\label{eq_ent in phase 2}
     S_R(A: B) = \lim_{m, n \rightarrow 1} \frac{1}{1-n} \log  \frac{\left\langle \sigma_{g_B}(-x_2) \sigma_{g_B g_A^{-1} }(x_1) \sigma_{g_B^{-1}}(x_2) \right\rangle_{\mathrm{CFT}^{\otimes mn}}}{\left(\left\langle \sigma_{g_m}(-x_2) \sigma_{g_m^{-1}}(x_2) \right\rangle_{\mathrm{CFT}^{\otimes m}}\right)^n} \,,
\end{align}
which is illustrated in Fig.~\ref{fig2}.
The three-point function in the numerator is computed as
\begin{equation}\label{ThreePointFunction}
    \left\langle \sigma_{g_B}(-x_2) \sigma_{g_B g_A^{-1}}(x_1) \sigma_{g_B^{-1}}(x_2) \right\rangle_{\mathrm{CFT}^{\otimes mn}} = \frac{C_{n, m}}{(x_2 - x_1)^{2h_{g_B g_A^{-1}}} (x_2 + x_1)^{2h_{g_B g_A^{-1}}} (2x_2)^{2(2h_{g_B} - h_{g_B g_A^{-1}})}}\,,
\end{equation}
The OPE coefficient is given by \cite{BasakKumar:2022stg, Dutta:2019gen}
\begin{align}\label{OPEcoefficent}
    C_{n, m}=e^{2(1-n) S_{\mathrm{bdy}}}(2 m)^{-4 h_{n}}, \;\;\; h_n:=\frac{ c}{24}\left(n-\frac{1}{n}\right)=\frac{h_{g_B g_A^{-1}}}{2} \,,
\end{align}
and this formula can be applied as $n\to1$ because $\sigma_{g_A}=\sigma_{g_B}$ at $n\to1$ \cite{Dutta:2019gen}.
Note that $(2 x_2)^{-2(2 h_{g_B})}$ in \eqref{ThreePointFunction} is cancelled by the denominator of equation (\ref{eq_ent in phase 2}). As a result, the reflected entropy for phase II \eqref{eq_ent in phase 2} is obtained as
\begin{align}\label{eq_SR phase 2}
    S_R(A:B)=\frac{c}{3}\log{\left(\frac{(x_2-x_1)(x_2+x_1)}{x_2\,\delta_x}\right)} \,,
\end{align}
where $\delta_x$ is the UV-cutoff. Equations \eqref{eq_SR phase 1} and \eqref{eq_SR phase 2} represent the reflected entropy of the two subregions $A=[0,x_1]$, $B=[x_1,x_2]$ in BCFT$_2$ where we employed the replica trick and the doubling trick. In the next section, we consider the same configurations and calculate the $m-1$ correction to the reflected entropies \eqref{eq_SR phase 1} and \eqref{eq_SR phase 2} under the assumption of a large central charge.

\subsection{$m-1$ correction to reflected entropy in BCFT$_2$}\label{sec_2.2}
In this section, we consider the $m-1$ correction of the Renyi reflected entropy for an adjacent configuration anchored to the boundary in BCFT$_2$.
When $m \neq 1$, there is a case in which the correlation function does not factorize. Therefore, we aim to calculate the reflected entropy that depends on $m$. If we consider the configuration of two adjacent intervals that is anchored to the boundary of BCFT$_2$, the Renyi reflected entropy is written as:
\begin{align}
    S_R^{(m)}(A:B)=\lim _{n \rightarrow 1} \frac{1}{1-n} \log \frac{\left\langle\sigma_{g_B g_A^{-1} }\left(x_1\right) \sigma_{g_B^{-1}}\left(x_2\right)\right\rangle_{BC F T^{\otimes m n}}}{\left(\left\langle\sigma_{g_m^{-1}}\left(x_2\right)\right\rangle_{BC F T^{\otimes m}}\right)^n}\,.
\end{align}
By employing the doubling trick, we obtain
\begin{align}\label{eq_Reflected entropy form 1}
 S_R^{(m)}(A:B)&=\lim_{n \rightarrow 1} \frac{1}{1-n} \log \frac{\left\langle\sigma_{g_B}\left(-x_2\right) \sigma_{g_A g_B^{-1} }\left(-x_1\right) \sigma_{g_B g_A^{-1} }\left(x_1\right) \sigma_{g_B^{-1}}\left(x_2\right)\right\rangle_{C F T^{\otimes m n}}}{\left(\left\langle\sigma_{g_m}\left(-x_2\right) \sigma_{g_m^{-1}}\left(x_2\right)\right\rangle_{C F T^{\otimes m}}\right)^n} \nn\\
 &\equiv \lim _{n \rightarrow 1} \frac{1}{1-n} \log{\frac{Z_{mn}}{(Z_m)^n}}\,,
\end{align}
where the partition function \eqref{eq_def partition function} for two adjacent intervals is given by
\begin{align}
     Z_{mn}:= \langle \sigma_{g_B}  ( &-x_2 ) \sigma_{g_A g_B^{-1}} (-x_1 ) \sigma_{g_B g_A^{-1}} ( x_1 ) \sigma_{g_B^{-1}} (x_2)  \rangle_{CFT^{\otimes m n}}, \quad Z_m:=Z_{m n}|_{n \rightarrow 1}. 
\end{align}
\subparagraph{(i) Phase I.}
For $m=1$, the four-point function (\ref{eq_Reflected entropy form 1}) factorizes into two-point functions. However, this factorization does not hold when $m\neq 1$. It becomes necessary to compute the four-point function (\ref{eq_Reflected entropy form 1}) for $m \neq 1$.

In CFT$_2$ with the Virasoro symmetry, the four-point function can be evaluated through the expansion of the conformal blocks, which are expressed as a sum over primary operators $\mathcal{O}_p$ with the conformal dimensions $h_p $ and $ \bar{h}_p$. This method is detailed in \cite{Ginsparg:1988ui, Hartman:2013mia,  Perlmutter:2015iya}.
\begin{equation}\label{ConformalBlockExpansion}
    \begin{aligned}
& \left\langle\sigma_{g_B}\left(-x_2\right) \sigma_{g_A g_B^{-1} }\left(-x_1\right) \sigma_{g_B g_A^{-1} }\left(x_1\right) \sigma_{g_B^{-1}}\left(x_2\right)\right\rangle_{C F T^{\otimes m n}} \\
= & \frac{1}{\left(2x_1\right)^{2(h_{g_B g_A^{-1}}+\bar{h}_{g_B g_A^{-1}})}\left(2x_2\right)^{2(h_{g_B}+\bar{h}_{g_B})}} \sum_p b_p \,\mathcal{F}\left(m n c, h_i, h_p, 1-z\right) \mathcal{F}\left(m n c, \bar{h}_i, \bar{h}_p, 1-\bar{z}\right)\,,
\end{aligned}
\end{equation}
where the cross-ratio is defined as
\begin{equation}\label{eq_CFT cross ratio}
    z=\frac{(x_2-x_1)^2}{(x_2+x_1)^2} \,,
\end{equation}
with its complex conjugate $\bar{z}$.\footnote{We are working with the real values $x_1$ and $x_2$, so $z=\bar{z}$.}
Additionally, $b_p$ is a product of the OPE coefficients, and $h_i$ and $\bar{h}_i$ represent the four conformal dimensions of the twist operators shown in equation (\ref{ConformalBlockExpansion}).  
By an analysis of the Liouville theory, the Virasoro conformal block, $\mathcal{F}$, is typically exponentiated as
\begin{equation}\label{eq_conformal block}
    \log \left[\mathcal{F}\left(m n c, h_i, h_p, 1-z\right)\right] \sim-\frac{m n c}{6} f\left(\epsilon_i, \epsilon_p, 1-z\right) \,,
\end{equation}
in the semi-classical limit
\begin{align}\label{eq_semi classical}
    m n c \rightarrow \infty, \quad \epsilon_i:=\frac{6 h_i}{m n c} \quad \text { and } \quad \epsilon_p:=\frac{6 h_p}{m n c} \quad \text { fixed }\,,
\end{align}
where we defined the rescaled conformal dimensions $\epsilon_i$ for each twist operators in equation (\ref{ConformalBlockExpansion}) on the $mn$ replica sheets.
Since we consider holographic CFTs, the identity block with $\epsilon_p=0$ becomes the dominant contribution in the conformal block expansion.\footnote{In our computation, we assume that the Virasoro identity block is dominant. In an orbifold CFT, it would be better to calculate using the orbifold block instead of the Virasoro block. Our CFT computation using the Virasoro block agrees one of the saddle solutions on the gravity side as shown in the next section. If we use the orbifold block instead of the Virasoro block, the CFT calculation in that case might be consistent with another saddle solution on the gravity side. It is an important future work to confirm this.} In this case, $b_p$ is simply given by $ b_p=1$.
Consequently, for the partition function $Z_{nm}$ in equation (\ref{eq_Reflected entropy form 1}) in the holomorphic part, we obtain
\begin{align}\label{eq_log Zmn phase 1}
    \log{Z_{mn}}=&-2 h_{g_B} \log{(2x_2)-2 h_{g_B g_A^{-1}} \log{(2x_1)}}-\frac{mnc}{6}f(\epsilon_i,0,1-z) \,.
\end{align}
For the partition function $Z_{m}$ in equation (\ref{eq_Reflected entropy form 1}), related to the holomorphic part, we get
\begin{equation}\label{eq_log Zm phase 1}
    n \log{Z_m}=-2 h_{g_B}\log{(2x_2)} \,.
\end{equation}
So far, we have not put the details of the conformal block $\mathcal{F}$. The exponent of the Virasoro identity block in the semi-classical limit at linear order in $\epsilon_{g_B g_A^{-1}}$, but working non-perturbatively in $\epsilon_{g_B}$, is outlined in \cite{Fitzpatrick:2014vua} as\footnote{The cross-ratio $z$ in our notation differs from the cross-ratio in \cite{Fitzpatrick:2014vua} by $z\leftrightarrow (1-z)$. We subtract $2\epsilon_{g_B g_A^{-1}}\log(1-z)$ since our definition of the Virasoro block does not include $(2x_1)^{-2h_{g_B g_A^{-1}}}$.}
\begin{align}\label{fallorder}
f(\epsilon_i,0,1-z)=&\;-2\epsilon_{g_B g_A^{-1}}\log \left(1-z\right)+2\epsilon_{g_B g_A^{-1}}\log\left(\frac{1-z^{\alpha_{g_B}}}{\alpha_{g_B}}\right)+\epsilon_{g_B g_A^{-1}}(1-\alpha_{g_B})\log z\,,
\end{align}
where $\alpha_{g_B}:=\;\sqrt{1-4\epsilon_{g_B}}$.
For the conformal dimensions for the twist operators on the $mn$ replica sheets \eqref{conformaldimension}, the rescaled conformal dimensions \eqref{eq_semi classical} behave in the $n-1$ expansion and the $m-1$ expansion respectively as:
\begin{align}
    &\epsilon_{g_{B}g_{A^{-1}}}\sim \frac{1}{m}(n-1)-\frac{3}{2m}(n-1)^2+\mathcal{O}\left((n-1)^3\right)\,,\label{eq_epsilon expansion1}\\ 
    &\epsilon_{g_B}\sim \frac{1}{2}(m-1)-\frac{3}{4}(m-1)^2+\mathcal{O}\left((m-1)^3\right)\,.\label{eq_epsilon expansion2}
\end{align}
Thus, the exponent of the conformal block up to the leading order in $\epsilon_{g_{B}g_{A^{-1}}}$ \eqref{fallorder} is in a valid regime since we are interested in the Renyi reflected entropy at $n\rightarrow 1$. Also, for $n\rightarrow 1$ and the large central charge, the $\epsilon_{g_B}$-expansion agrees with the $m-1$ expansion up to their linear orders.
Putting everything together, the Reny reflected entropy \eqref{eq_Reflected entropy form 1} becomes
\begin{align}\label{eq_SR all order m 1}
    S_R^{(m)}(A:B)=\frac{c}{3}\log \left(\frac{2 x_1}{\delta_x}\right)+\frac{c}{3}\log\left[\frac{m\sqrt{z}}{1-z}\left(z^{-\frac{1}{2m}}-z^{\frac{1}{2m}}\right) \right]\,,
\end{align}
where only the holomorphic part was taken, accounting for the chiral CFT. Expanding the Renyi reflected entropy \eqref{eq_SR all order m 1} in orders of $m-1$, we obtain
\begin{equation}\label{BCFTPH1}
    S_R^{(m)}(A:B)=\frac{c}{3}\log \left(\frac{2 x_1}{\delta_x}\right)+\frac{c}{6}\left(2+\frac{1+z}{1-z}\log{z}\right)(m-1)+\mathcal{O}\left((m-1)^2\right) \,.
\end{equation}
The leading order term in equation \eqref{BCFTPH1} aligns well with the reflected entropy for phase I, as given in \eqref{eq_SR phase 1}. The results in equations \eqref{eq_SR all order m 1} and \eqref{BCFTPH1} demonstrate how the $m-1$ correction to the reflected entropy is captured through the operator product expansion involving the semi-classical Virasoro block. 
In particular, the second term in equation \eqref{eq_SR all order m 1} explicitly represents the non-perturbative dependence of the Renyi reflected entropy on $m$, while its expansion in equation \eqref{BCFTPH1} reveals the leading order correction in small $m-1$ limit. It is important to note that the Renyi reflected entropy depends explicitly on the cross-ratio $z$. As will be shown in the analysis of phase II, this $z$-dependence gives rise to a phase transition in the Renyi reflected entropy.

\subparagraph{(ii) Phase II.}
In phase II, as in the discussion at the previous subsection, we also assume that the numerator of the Renyi reflected entropy (\ref{eq_Reflected entropy form 1}) can be evaluated using equation (\ref{phaseII}) even for $m\ne1$ because the value of $m$ is not important for treating either $\sigma_{g_A g_B^{-1} }(-x_1)$ or $\sigma_{g_B g_A^{-1} }(x_1)$ as the identity operator at $n\to1$.
The Renyi reflected entropy for phase II is then expressed as
\begin{equation}\label{eq_reflected ent phase II}
     S_R^{(m)}(A: B)=\lim _{n \rightarrow 1} \frac{1}{1-n} \log {\frac{\left\langle\sigma_{g_B}\left(-x_2\right)\sigma_{g_B g_A^{-1} }\left(x_1\right) \sigma_{g_B^{-1}}\left(x_2\right) \right\rangle_{\mathrm{CFT}^{\otimes m n}} }{\left(\left\langle\sigma_{g_m}\left(-x_2\right) \sigma_{g_m^{-1}}\left(x_2\right)\right\rangle_{C F T^{\otimes m}}\right)^n}} \,,
\end{equation}
where the three-point function is outlined in equation (\ref{ThreePointFunction}).
Consequently, the partition function $Z_{mn}$ in equation (\ref{eq_Reflected entropy form 1}), we obtain
\begin{equation}
    \log{Z_{mn}}=-2 h_{g_B g_A^{-1}} \log \left(\frac{\left(x_2-x_1\right)\left(x_2+x_1\right)}{2 x_2}\right)-4h_{g_B}\log{(2x_2)}+\log C_{n, m} \,,
\end{equation}
where the OPE coefficient is given by \eqref{OPEcoefficent}.
For the partition function $Z_m$ in equation (\ref{eq_Reflected entropy form 1}), we have
\begin{equation}
    n\log{Z_m}=-4h_{g_B}\log{(2x_2)} \,.
\end{equation}
Therefore, the Renyi reflected entropy is calculated as
\begin{equation}
\begin{aligned}\label{REBCFTPH2}
S_R^{(m)}(A:B)=\frac{c}{3} \log{\left(\frac{(x_2-x_1)(x_2+x_1)}{x_2\,\delta_x}\right)} +\frac{c}{3}\log{m}\,.
\end{aligned}
\end{equation}
Expanding the Renyi reflected entropy \eqref{REBCFTPH2} in orders of the $m-1$, we obtain
\begin{equation}
\begin{aligned}\label{eq_REBCFTPH2 order}
S_R^{(m)}(A:B)=\frac{c}{3} \log{\left(\frac{(x_2-x_1)(x_2+x_1)}{x_2\,\delta_x}\right)} +\frac{c}{3}(m-1)+\mathcal{O}\left((m-1)^2\right).
\end{aligned}
\end{equation}
The leading order term in equation \eqref{eq_REBCFTPH2 order} reproduces the reflected entropy for phase II as given in \eqref{eq_SR phase 2}. Equations \eqref{REBCFTPH2} and \eqref{eq_REBCFTPH2 order} illustrate how the $m-1$ correction to the reflected entropy appears in phase II. The $\log m$ term in \eqref{REBCFTPH2}, originating from the OPE coefficient in \eqref{OPEcoefficent}, captures the non-perturbative $m$-dependence. The Renyi reflected entropy at a given cross-ratio $z$ is determined by the minimum of equations \eqref{eq_SR all order m 1} and \eqref{REBCFTPH2}. This behavior demonstrates how the presence of the boundary in AdS/BCFT modifies the entanglement structure, resulting in a phase transition in the Renyi reflected entropy—a boundary-induced effect that is absent in the standard AdS/CFT case.

In the next section, we will calculate the entanglement wedge cross section. Considering the duality between the $m-1$ correction to the reflected entropy in CFT$_2$ and the backreaction effect on the entanglement wedge cross section \cite{Jeong:2019xdr}, we aim to establish this duality even in the presence of the boundary of BCFT$_2$ and the end of the world brane in AdS$_3$.

%
\section{Entanglement wedge cross section with the backreaction from cosmic branes}\label{sec_3}
In this section, our goal is to calculate the entanglement wedge cross section for two adjacent intervals, where one end point of an interval is on the boundary of the BCFT. Specifically, we are interested in exploring the entanglement wedge cross section in cases that include the backreaction of the cosmic brane. Through these calculations, we aim to illustrate that the $m-1$ corrections to the reflected entropy in BCFT$_2$ align with the modifications to the entanglement wedge cross section in AdS$_3$ gravity, induced by the end of the world brane.
Specifically, we are interested in the validity of the following relation in the context of AdS$_3$/BCFT$_2$ \cite{Dutta:2019gen}
\begin{equation}\label{eq_reflected and EWCS}
    S_R^{(m)}(A: B)=2 E_{W}^{(m)}(A: B)+\mathcal{O}(G_N^0)\,.
\end{equation}
We evaluate the reflected entropy on the left-hand side of \eqref{eq_reflected and EWCS} up to the leading order in $1/c$ expansion, which corresponds to the small $G_N$ expansion on the bulk side.
$E_{W}^{(m)}$ on the right-hand side is the entanglement wedge cross section with $m-1$ correction, which will be defined as a minimal area splitting the entanglement wedge divided by $4 G_N$ where $G_N$ is the gravitational constant.
This alignment supports the proposed duality between the classical geometry and its dual CFT.

To facilitate this exploration, we employ the method of computing the backreaction of the cosmic brane \cite{Dong:2016fnf}. Before engaging in calculations that involve backreaction effects, we first review the concept of the entanglement wedge cross section and describe the procedure for its calculation in configurations involving two adjacent intervals within the pure AdS$_3$ geometry. Furthermore, we investigate the calculation of the entanglement wedge cross section by introducing the end of the world brane, which serves as the gravity dual representation of BCFT. The inclusion of the end of the world brane expands our ability to examine a broader spectrum of phases.
Subsequently, our analysis will focus on computing the entanglement wedge cross section with a small backreaction, specifically for intervals that originate at the point where the boundary of BCFT is located.

\subsection{Entanglement wedge cross section in AdS$_3$/BCFT$_2$}\label{sec_3.1.2}

\paragraph{Entanglement wedge cross section.} 
Let us start by considering the entanglement wedge cross section in the context of a pure AdS$_3$ geometry, without the influence of backreaction. The unperturbed metric of this spacetime is represented as:
\begin{align}\label{Poincare}
    ds^2=\frac{d\xi^2+dt^2+dx^2}{\xi^2} \,,
\end{align}
where the AdS radius is set to unity, and $t$ is the Euclidean time. The boundary of this pure AdS$_3$ is located at $\xi\rightarrow 0$. We consider two subregions $A=[x_1,x_2]$ and $B=[x_3,x_4]$ with $x_1<x_2<x_3<x_4$ on the boundary at a fixed time $t=0$. According to the Ryu-Takayanagi conjecture, there are two potential configurations for the minimal surface $\Gamma^{min}_{AB}$ associated with the union of $A$ and $B$: 
one is the connected configuration, and the other one is the disconnected configuration.\footnote{In 2-dimensional CFT, configuration of the RT surface is determined by the cross-ratio $z=\frac{\left(x_2-x_1\right)\left(x_4-x_3\right)}{\left(x_3-x_1\right)\left(x_4-x_2\right)}$. For $0<z<1/2$, the disconnected configuration is dominant (minimal). For $1/2<z<1$, the connected configuration is dominant (minimal).}
The boundary of the entanglement wedge, denoted $M_{AB}$, is defined by $\partial M_{AB}=A\cup B\cup \Gamma^{min}_{AB}$. Within this wedge, the entanglement wedge cross section $E_W(A:B)$ is defined as the codimension-2 minimal surface that divides the entanglement wedge $M_{AB}$ into two codimension-1 bulk subregions $M_A$ and $M_B$ such that $M_A$ includes the boundary subregion $A$, and $M_B$ includes the boundary subregion $B$.

In cases where the Ryu-Takayanagi (RT) surface is disconnected, $M_A$ and $M_B$ are inherently separate, leading to the absence of an entanglement wedge cross section. Conversely, in the connected configuration, a codimension-2 surface $\Sigma_{AB}$ exist, effectively dividing $M_{AB}$ into $M_A$ and $M_B$. Here, the entanglement wedge cross section is quantitatively defined by the minimization of the area $\Sigma_{AB}$:
\begin{align}
    E_W(A: B):=\text{min}\left[\frac{\operatorname{Area}\left(\Sigma_{A B}\right)}{4 G_N}\right] \,,
\end{align}
where $G_N$ represents the gravitational constant. 
In \cite{Dutta:2019gen}, it was shown that twice the entanglement wedge cross section in the bulk spacetime is equivalent to the reflected entropy for the corresponding subregions on the boundary of AdS spacetime
\begin{align}
    S_R(A: B)=2 E_W(A: B)\,.
\end{align}

\paragraph{End of the world brane.}
A simple model of AdS/BCFT is described by an asymptotically AdS spacetime with an end of the world brane \cite{Karch:2000gx, Takayanagi:2011zk, Fujita:2011fp}. The end of the world brane extends from the boundary of BCFT. We choose the Neumann boundary condition on the end of the world brane for probing the dynamics of brane.
The existence of the end of the world brane in AdS spacetime corresponds to the boundary degrees of freedom such as $\mathcal{C}_{\mathcal{O}}$ in (\ref{tpBCFT}). For example, the non-vanishing one-point function in the BCFT \eqref{tpBCFT} can be represented as a minimal geodesic curve reflected by the end of the world brane in the holographic setup \cite{Park:2024pkt}. In a similar fashion, one can see how the non-trivial phases of reflected entropy in BCFT can be visualized as a richer set of geometric configurations in the AdS/BCFT setup. When the end of the world brane exists, the minimal surface $\Gamma^{min}_{AB}$ can anchor at the end of the world brane, and the boundary of the entanglement wedge $\partial M_{AB}$ can include a part of the end of the world brane. Additionally, the entanglement wedge cross section $E_W(A:B)$ can also anchor at the end of the world brane, as well as at the minimal surface $\Gamma^{min}_{AB}$. Therefore, the presence of the end of the world brane makes a rich phase structure of geometric configurations of $E_W(A:B)$.

\paragraph{Entanglement wedge cross section in AdS/BCFT.}
First, we compute the entanglement wedge cross section for two adjacent intervals on a fixed time slice $t=0$. These intervals, of particular interest in our study, are given as $A=[0,x_1]$ and $B=[x_1,x_2]$. In the context of a BCFT, there is only one possible configuration for the entanglement wedge of the union $A \cup B$. This entanglement wedge is bounded by the cosmic brane that is anchored at the boundary point $x_2$ and extends to the end of the world brane. Within this framework, the entanglement wedge cross section can manifest in two distinct phases: (i) The cross section anchored at the end of the world brane (ii) The cross section anchored at the cosmic brane. Each phase corresponds to different configurations of the minimal surfaces within the entanglement wedge, reflecting various quantum entanglement properties between the intervals $A$ and $B$.

\begin{figure}
\begin{center}
\begin{tabular}{cc}
\setlength{\unitlength}{1cm}
\hspace{0.1cm}
\includegraphics[width=7cm]{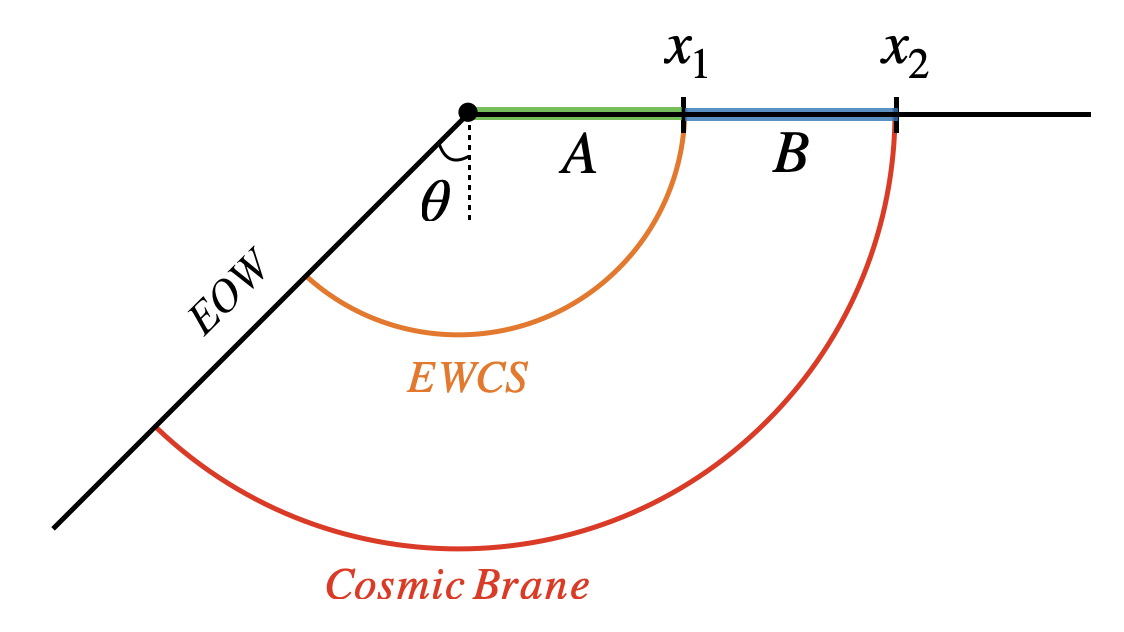}
\includegraphics[width=7cm]{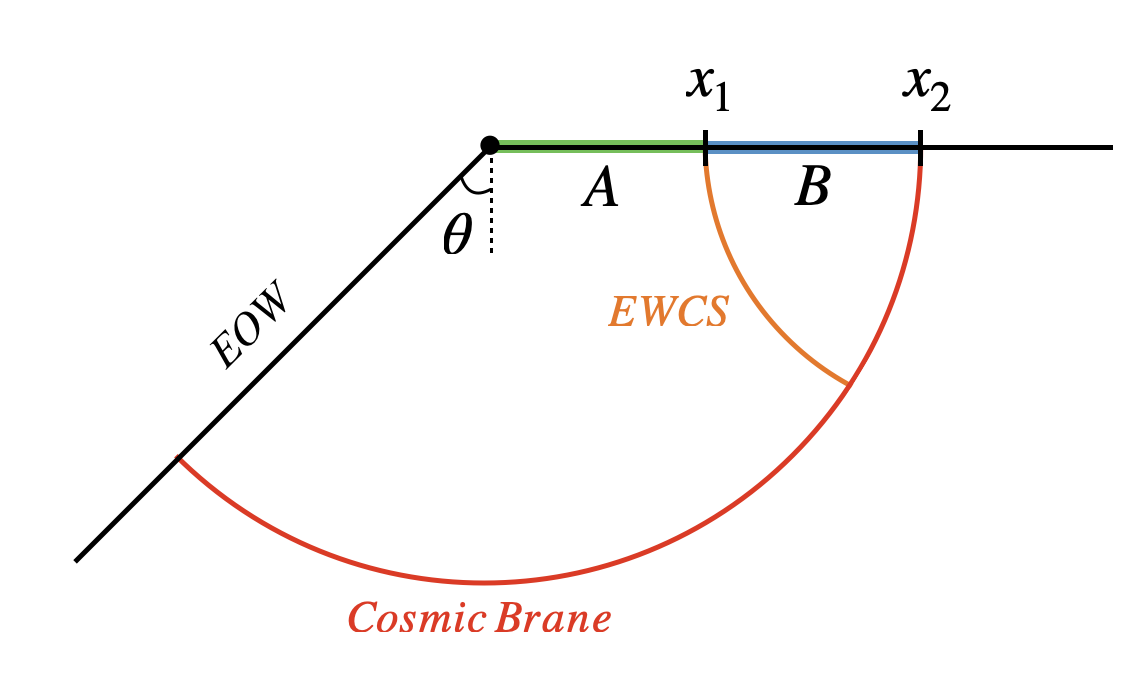}
\qquad
\end{tabular}
\end{center}
\caption{Spacetime diagram of phase 1 (Left) and phase 2 (Right) for adjacent intervals in contact with the boundary of the BCFT. The horizontal line is the $x$ direction and the vertical line is the $z$ direction on a constant $t$ slice. The red surface indicates the cosmic brane (for $m\neq 1$), which becomes the RT minimal surface for the boundary subregion $A\cup B$ as $m\rightarrow 1$. The orange surface indicates the entanglement wedge cross section with backreaction from a cosmic brane. The profile of the end of the world brane is parametrized by the angle $\theta$, which is related to the boundary entropy $S_{\mathrm{bdy}}$.}
\label{fig6}
\end{figure}

\subparagraph{(i) Phase I.} As depicted in Fig.~\ref{fig6}, this phase of the entanglement wedge cross section is anchored to the end of the world brane. In this configuration, the entanglement wedge cross section is determined by minimizing the area of a surface that begins at $x_1$ and extends to  a point on the end of the world brane within Poincare geometry. The calculation for this cross section is expressed as 
\begin{align}\label{eq_zeroth EW simple 1}
    E_W(A: B)=\frac{1}{4 G_N}\log{\left(\frac{2x_1}{\delta_x}\right)} \,,
\end{align}
where $\delta_x$ is the UV-cutoff, and we assume $\theta=0$, implying that the boundary entropy $S_{bdy}$ is zero.

\subparagraph{(ii) Phase II.} In this phase, the entanglement wedge cross section initiates at $x_1$ and terminates at a point on the cosmic brane. Here, the entanglement wedge cross section is determined by calculating  the minimal area of the surface connecting  $x_1$ to the cosmic brane within Poincare geometry. This formulation is given by
\begin{align}\label{eq_zeroth EW simple 2}
    E_W(A: B) =\frac{1}{4G_N} \log{\left(\frac{(x_2-x_1)(x_2+x_1)}{x_2 \delta_x}\right)}  \,,
\end{align}
where $\delta_x$ is the UV-cutoff.

\subsection{Backreaction on entanglement wedge cross section}\label{sec_EWCS with backreaction}\label{sec_3.2}
In this section, we compute the entanglement wedge cross section for the two adjacent intervals $A=[0,x_1]$ and $B=[x_1,x_2]$, considering the backreaction of the cosmic brane. For simplicity, we consider only the case the end of the world brane is orthogonal to the boundary of AdS spacetime, allowing us to ignore its backreaction effects, which is equivalent to the vanishing boundary entropy $S_{bdy}=0$ in BCFT.

\paragraph{Cosmic brane.} 
Cosmic brane is an extension of the RT minimal surface. Specifically, a cosmic brane serves as the gravity dual of the Renyi entropy \cite{Dong:2016fnf}, whereas the RT minimal surface, without backreaction, corresponds to the gravity dual of the entanglement entropy. The backreaction is parametrized by $m$, which is associated with the tension of the cosmic brane $T_m=\frac{m-1}{4 m G_N}$. When $m=1$, the tension of the cosmic brane vanishes, indicating that there is no backreaction and the cosmic brane reverts to the standard RT surface. 

\paragraph{Entanglement wedge cross section with backreaction.}
In this discussion, we analyze the impact of backreaction on the entanglement wedge cross section, involving the influence of the cosmic brane. Consequently, we replace the traditional RT surface with one affected by the cosmic brane.
The cosmic brane  perturbs the geometry.
Then, a codimension-2 surface that divides the entanglement wedge receives dependency on $m$:
\begin{align}
    E_{W}^{(m)}(A: B):=\text{min}\left[\frac{\operatorname{Area}(\Sigma_{A B}^{(m)})}{4 G_N}\right] \,.
\end{align}
In this formulation, We replace $\Sigma_{AB}$ with $\Sigma_{AB}^{(m)}$ by promoting the codimension-2 surface to depend on the parameter $m$.
In the bulk perspective, the parameter $m$ directly corresponds to the tension of the cosmic brane. Note that the tension of the codimension-2 surface $\Sigma_{A B}^{(m)}$ is zero, but $\Sigma_{A B}^{(m)}$ depends on $m$ via the nonzero tension of the cosmic brane. From the boundary perspective, it is associated with the replica index $m$.

To compute the contribution from the backreaction of the single cosmic brane, it is useful to employ the hyperbolic black hole geometry, following the calculations detailed in \cite{Hung:2011nu, Dong:2016fnf, Jeong:2019xdr}. The underlying concept of the calculation is as follows: The entanglement entropy of a spherical subregion in the vacuum state of the CFT in $d$-dimensional Minkowski space can be equated with the thermal entropy of the CFT on $d$-dimensional hyperbolic geometry, where the thermal entropy of the $d$ dimensional CFT can be evaluated by the area of the $(d+1)$-dimensional hyperbolic black holes in the holographic argument \cite{Casini:2011kv}. Using this approach, the Renyi entropy for a spherical subregion was calculated in \cite{Hung:2011nu} by considering the thermal entropy of the CFT on the unit hyperboloid, with the inverse temperature $\beta=2\pi m$, where $m$ is the Renyi index.  Building on this insight, the author in \cite{Dong:2016fnf} demonstrated that the gravity dual of the Renyi entropy is the cosmic brane, resulting in the same conical singularity as that of the hyperbolic black hole geometry with inverse temperature $\beta = 2\pi m$. This establishes our understanding of the geometry associated with the cosmic brane. Therefore, by computing the entanglement wedge cross section within this geometry, we would be able to elucidate the impact of the cosmic brane's backreaction on the entanglement wedge cross section. This reasoning underpins our approach to calculating the entanglement wedge cross section within the hyperbolic black hole geometry, which closely resembles the geometry modified by the backreaction of the single cosmic brane.

The geometry involving the cosmic brane can be obtained through an appropriate coordinate transformation from the Poincare geometry, and it is described by the metric
\begin{align}\label{BHgeometry}
   \mathrm{d} s^2=\frac{d r^2}{r^2-r_h^2}+\left(r^2-r_h^2\right) d \tau^2+r^2 d \rho^2 \,,
\end{align}
where $r_h=\frac{1}{m}$ represents the horizon with an inverse temperature $\beta=\frac{2\pi}{r_h}$, and $\tau$ is the Euclidean time. The cosmic brane envelops the horizon in the hyperbolic black hole geometry. The hyperbolic black hole \eqref{BHgeometry} describes effectively the bulk geometry with the cosmic brane, as the horizon and the brane share the same  conical singularity \cite{Dong:2016fnf, Hung:2011nu, Casini:2011kv}.

In the case of two adjacent intervals where one end of one interval is located at the origin, there is only one RT surface that constitutes the entanglement wedge. We treat the RT surface as the cosmic brane with nonzero tension to incorporate the backreaction effect.

For two intervals $A=[0,x_1]$ and $B=[x_1,x_2]$, a global conformal transformation ensures the invariance of the cross ratio under the transformation \cite{Jeong:2019xdr, Dong:2016fnf}
\begin{equation}\label{eq_gravity cross ratio}
z = \frac{(x_2-x_1)^2}{(x_2+x_1)^2} = \frac{(1-R_0)^2}{(1+R_0)^2}, \quad R_0=\frac{x_1}{x_2}.
\end{equation}
This transformation modifies the points as follows:
\begin{equation}\label{eq_conf transf 1}
-x_2 \rightarrow -1, \quad -x_1 \rightarrow -R_0, \quad x_1 \rightarrow R_0, \quad x_2 \rightarrow 1 \,,
\end{equation}
and the interval is redefined as $A=[0,R_0]$ and $B=[R_0,1]$. Under the bulk coordinate transformation that converts the bulk geometry with the cosmic brane into the hyperbolic black hole geometry, the boundary coordinates transform as \cite{Jeong:2019xdr, Dong:2016fnf}
\begin{equation}\label{conformal transformation}
\tan \tau=\frac{2 t}{1-t^2-x^2}, \quad \tanh \rho=\frac{2 x}{1+t^2+x^2}.
\end{equation}
Following this transformation,
the intervals are defined as
\begin{equation}
\begin{array}{llllll}
A: & 0 \leq x \leq R_0 & (t=0) & \quad\rightarrow & \quad 0 \leq \rho \leq \rho_0 & (\tau=0) \\
B: & R_0 \leq x \leq 1 & (t=0) & \quad\rightarrow & \quad\rho_0 \leq \rho \leq \infty & (\tau=0) \,,
\end{array}
\end{equation}
where the inverse transformation is given by
\begin{equation}\label{inversetransformation 1}
    \rho_0=\tanh^{-1}{\left(\frac{2 R_0}{1+R_0^2}\right)}=-\frac{1}{2} \log z \,.
\end{equation}

Although the cross ratio remains invariant under the conformal transformation, the UV cutoff scale changes.
From the perspective of the holographic renormalization group flow, the UV cutoff is related to the energy scale of the theory. From the bulk perspective, we need to introduce the UV cutoff around $x_1$ since the entanglement wedge cross section is anchored at $x_1$. Thus, the UV cutoff given in the Poincare patch of pure AdS changes due to the conformal transformations (\ref{eq_conf transf 1}) and (\ref{conformal transformation}). For this reason, $\delta_\rho$ in the $(\tau, \rho)$ coordinates depends on the end points of the intervals. We provide a prescription for the relation between $\delta_x$ and $\delta_\rho$ as follows.

Let us introduce a point $x^{\prime}$ near $x_1$ in order to derive $\delta_\rho$ in terms of the endpoints in the $(t,x)$ coordinates and $\delta_x$.
\begin{equation}
    x^{\prime}=x_1+\delta_x \,.
\end{equation}
Assume $x^{\prime}$ transforms into $R^{\prime}=\frac{x^{\prime}}{x_2}=R_0+\delta_R$ due to the global conformal transformation from (\ref{eq_conf transf 1}). Then, after the conformal transformation \eqref{conformal transformation}, $R^{\prime}$ becomes
\begin{align}\label{eq_cutoff transf}
\begin{split}
   \rho^{\prime}&=\tanh^{-1}{\left(\frac{2 (R_0+\delta_R)}{1+(R_0+\delta_R)^2}\right)} \\
   &\simeq \rho_0+\delta_\rho +\mathcal{O}(\delta_x^2) \,.
\end{split}
\end{align}
By equating the first line and the second line, one can establish the relation between the cutoffs in $(\tau,\rho)$ and $(t,x)$ coordinates as follows:
\begin{equation}\label{inversetransformation}
    \delta_\rho=\frac{2 x_2 \delta_x}{(x_2^2-x_1^2)}.
\end{equation}
Here, $\delta_x$ represents the UV cutoff in $(t,x)$ coordinates.

\subparagraph{(i) Phase I.} 
In this case, the entanglement wedge cross section terminates at the end of the world brane. The area of the minimal surface connecting the boundary point $\rho_0$ and the end of the world brane is then given by \cite{Takayanagi:2017knl}
\begin{equation}\label{eq_phase 1 area}
    \text{Area}=\log{\left[\frac{\beta}{\pi \delta_\rho}\sinh{\left(\frac{2\pi \rho_0}{\beta}\right)}\right]}=\log\left[\frac{2m}{\delta_\rho}\sinh{\left(\frac{\rho_0}{m}\right)}\right]\,,
\end{equation}
where $\beta$ is the inverse temperature in the hyperbolic black hole geometry, and we used $\beta=2\pi m$ in the second equality. Here, $\delta_\rho$ is the UV cutoff in the $(\tau,\rho)$ coordinates.
Utilizing the inverse transformations \eqref{inversetransformation 1} and \eqref{inversetransformation} to the $(t,x)$ coordinates and the cross-ratio \eqref{eq_gravity cross ratio}, the entanglement wedge cross section with backreaction is derived as
\begin{eqnarray}\label{eq_EWCS phase 1}
    E_{W}^{(m)}(A: B)&=&\frac{1}{4 G_N}\log{\left(\frac{2x_1}{\delta_x}\right)}+\frac{1}{4 G_N}\log\left[\frac{m\sqrt{z}}{1-z}\left(z^{-\frac{1}{2m}}-z^{\frac{1}{2m}}\right) \right]\\
    &=&\frac{1}{2}S_R^{(m)}(A:B) \,,
\end{eqnarray}
where the Renyi reflected entropy \eqref{eq_SR all order m 1} in phase I and the Brown-Henneaux relation $c=\frac{3}{2G_N}$ \cite{Brown:1986nw} are used in the second line. The $m\rightarrow 1$ limit of the result aligns well with equation \eqref{eq_zeroth EW simple 1}. Notably, since this configuration receives contribution only from the single cosmic brane, the result in equation (\ref{eq_EWCS phase 1}) remains valid at all orders in $m-1$. This relation between the entanglement wedge cross section with backreaction and the Renyi reflected entropy in phase I is in agreement with \eqref{eq_reflected and EWCS}.

Equation \eqref{eq_EWCS phase 1} illustrates how the $m-1$ correction to the entanglement wedge cross section captures the effect of cosmic brane backreaction in the bulk, specifically when the entanglement wedge terminates on the end-of-the-world brane. The second term, which explicitly depends on $m$, arises from the deformation of the bulk geometry induced by the brane tension. To express this correction in terms of the cosmic brane tension $T_m = \frac{m - 1}{4 m G_N}$, we rewrite the second term in \eqref{eq_EWCS phase 1} as
\begin{align}
    \frac{1}{4G_N} \log\left[\frac{z^{2 G_N T_m} - z^{1 - 2 G_N T_m}}{(1 - z)(1 - 4 G_N T_m)}\right].
\end{align}
In the limit $T_m\to 0$, this correction vanishes, and we recover the entanglement wedge cross section in the absence of backreaction. Since $0\leq T_m <\frac{1}{4 G_N}$ for $m \in [1, \infty)$, the correction term is always negative within this range, indicating that the backreaction of the cosmic brane reduces the area of the entanglement wedge cross section.

\subparagraph{(ii) Phase II.} 
In this case, we consider the entanglement wedge cross section terminates on the horizon. The minimal area of the surface connecting the boundary point $x_2$ and the horizon of the hyperbolic black hole is given by \cite{Takayanagi:2017knl}
\begin{equation}\label{eq_phase 2 m correction}
    \text{Area} =\log{\left(\frac{\beta }{\pi \delta_\rho}\right)} = \frac{1}{4G_N}\log \left( {\frac{2}{\delta_\rho}} \right) +\frac{1}{4G_N}\log{m}\,,
\end{equation}
where we used $\beta=2\pi m$ in the second equality.
Although the expression appears independent on the boundary points, $\delta_\rho$ contains all the information about the boundary interval.
Utilizing the inverse transformation of the cutoff \eqref{inversetransformation}, we derive the entanglement wedge cross section with backreaction as follows:
\begin{align}\label{EWCSphase2}
    E_{W}^{(m)}(A: B) &=\frac{1}{4G_N} \log{\left(\frac{(x_2-x_1)(x_2+x_1)}{x_2 \delta_x}\right)} +\frac{1}{4G_N}\log{m}\\
    &=\frac{1}{2}S_R^{(m)}(A:B) \,,
\end{align}
where we used the Renyi reflected entropy \eqref{REBCFTPH2} in phase II and the Brown-Henneaux relation in the second line.
The first term in the first line corresponds to the entanglement wedge cross section \eqref{eq_zeroth EW simple 2} in Poincare geometry. As in phase I, the entanglement wedge cross section \eqref{EWCSphase2} involving the backreaction is valid at all orders in $m-1$. This relation between the entanglement wedge cross section with backreaction and the Renyi reflected entropy aligns with \eqref{eq_reflected and EWCS}.

Equation~\eqref{EWCSphase2} captures the $m-1$ correction to the entanglement wedge cross section due to the backreaction of a cosmic brane with tension \( T_m = \frac{m - 1}{4m G_N} \). The \(\log m\) term reflects the conical defect induced by the brane, and can be rewritten as \(\log(1 - 4 G_N T_m)^{-1}\), making its dependence on the brane tension explicit. As \( T_m \to 0 \), the correction vanishes, and we recover the result without backreaction. In contrast, as $m \to \infty$, the brane tension approaches $\frac{1}{4 G_N}$, causing the correction term—and hence the entanglement wedge cross section—to diverge. The reason for this divergence is that the place where the cosmic brane at $m\to \infty$ is located corresponds to the horizon at $r=r_h=\frac{1}{m}\to0$ in the metric (\ref{BHgeometry}). In phase II, the entanglement wedge cross section is anchored to the cosmic brane, leading to a divergent area because the distance to the horizon at $r=r_h=\frac{1}{m}\to0$ is divergent. In contrast, in phase I, the entanglement wedge cross section does not touch the cosmic brane as seen in Fig.~\ref{fig6}, so the area remains finite.

Importantly, the consistency between the entanglement wedge cross sections in equations \eqref{eq_EWCS phase 1} and \eqref{EWCSphase2} and the Renyi reflected entropies in equations \eqref{eq_SR all order m 1} and \eqref{REBCFTPH2} further solidifies the duality between boundary CFT calculations and bulk gravitational descriptions. This confirms that the $m-1$ expansion in the CFT precisely encodes the gravitational backreaction in the bulk.
%

\section{More on entanglement wedge cross section in AdS$_3$/BCFT$_2$}
\subsection{General adjacent intervals}\label{sec 4.1}

In this section, we compute the entanglement wedge cross section of two adjacent intervals, $A=[x_1,x_2]$ and $B=[x_2,x_3]$. By setting the distance between the boundary of the BCFT and interval A to a nonzero, $x_1\ne0$, we can explore additional possible phases.

\begin{figure}
\begin{center}
\begin{tabular}{cc}
\includegraphics[width=5cm]{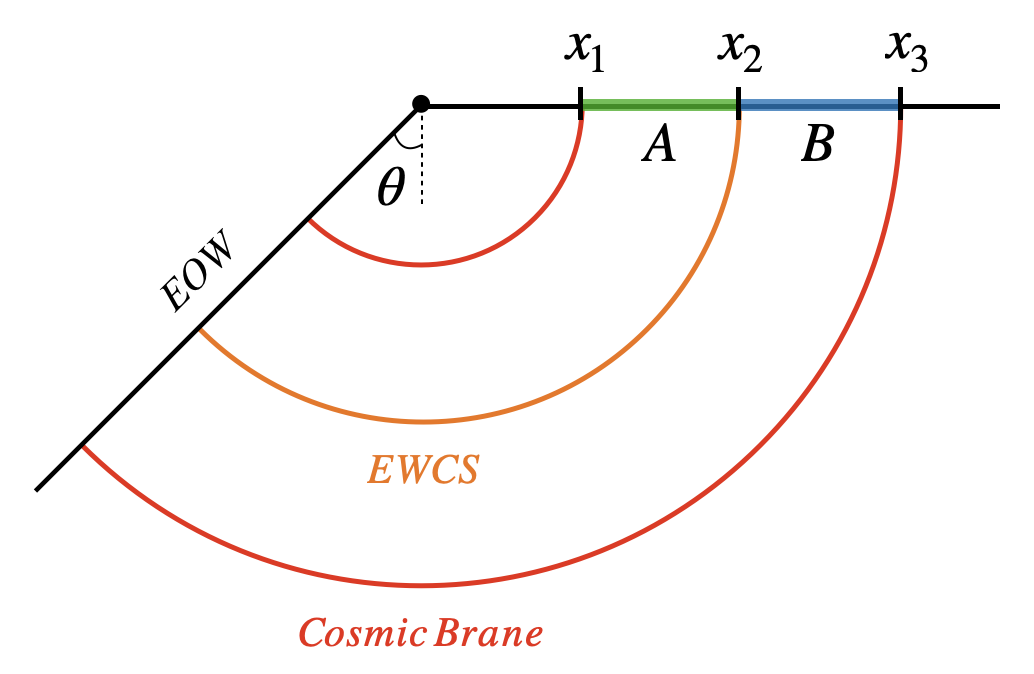}
\includegraphics[width=5cm]{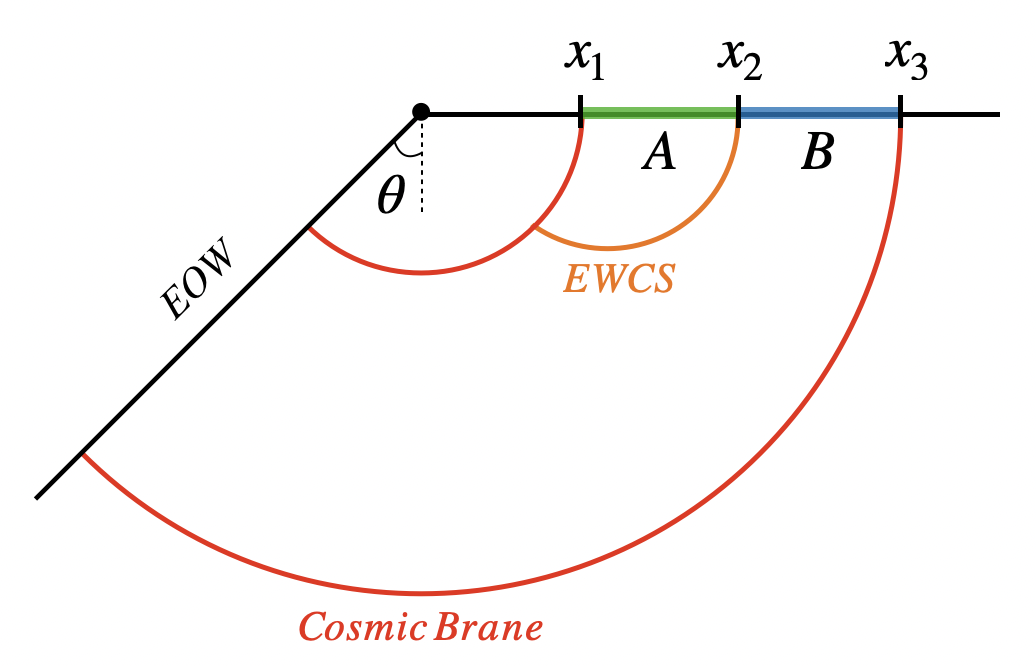}
\includegraphics[width=5cm]{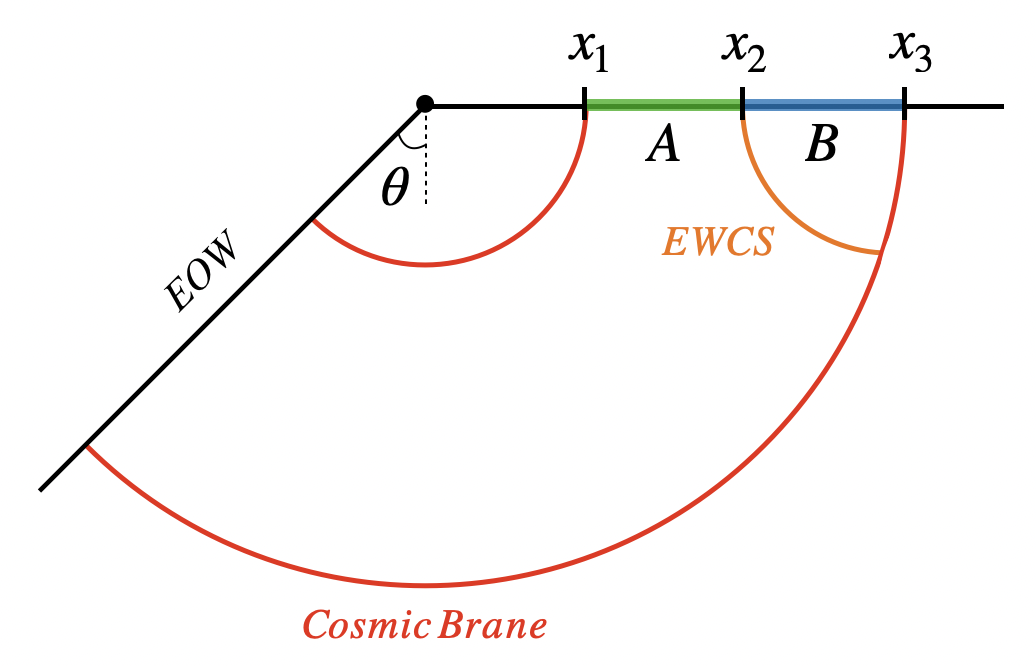}
\qquad
\end{tabular}
\end{center}
\caption{Spacetime diagram of phase 1 (Left), phase 2 (Middle), phase 3 (Right) in the disconnected Ryu-Takayanagi configuration for the general adjacent intervals.}
\label{fig7}
\end{figure}

\begin{figure}[ ]
    \centering
\includegraphics[width=0.46\textwidth]{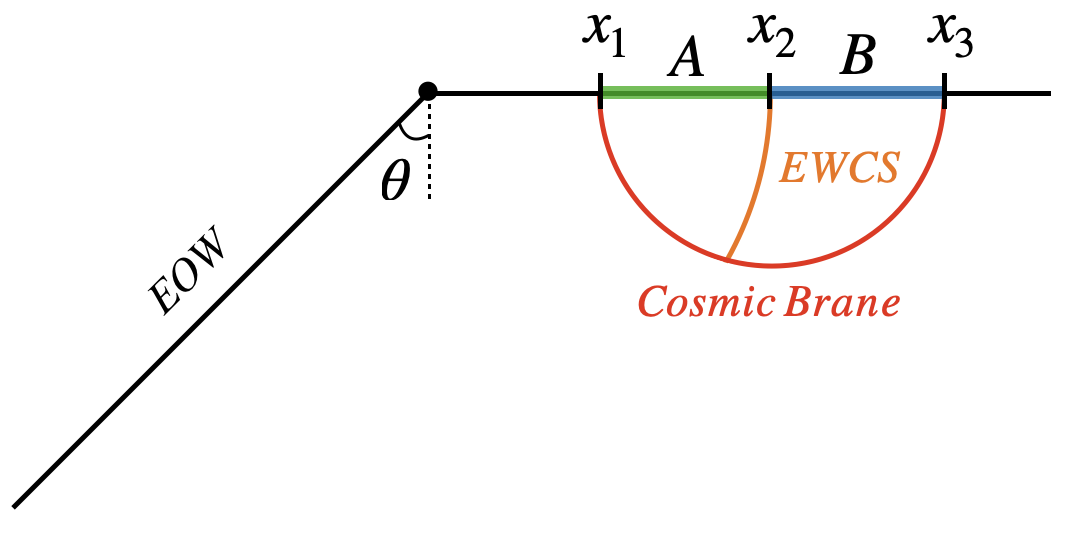}
    \caption{Spacetime diagram in the connected Ryu-Takayanagi configuration for the general adjacent intervals.}
    \label{fig_GeneralEWCS_Connected}
\end{figure}

For the adjacent two intervals that are not attached to the boundary of BCFT, there are two possible configurations of the Ryu-Takayanagi (RT) surface of the subregion $A\cup B$: the connected configuration and the disconnected configuration.  In the disconnected configuration, each endpoint of the subregion $A\cup B$ is anchored to the end of the world brane, creating two disconnected RT surfaces (Fig.~\ref{fig7}). In the connected configuration, two end points of the subregion $A\cup B$ are connected through a single RT surface (Fig.~\ref{fig_GeneralEWCS_Connected}).

Before moving on to calculations, the cross-ratios are first defined in the general adjacent intervals case.
By a global conformal transformation ensuring that the cross ratios remain invariant, we have
\begin{align}
\begin{split}
    z_+ &=\frac{(x_3-x_2)^2}{(x_3+x_2)^2} = \frac{(1-R_+)^2}{(1+R_+)^2} \,, \quad R_+ =\frac{x_2}{x_3} \,,\\
    z_0 &=\frac{(x_3-x_1)^2}{(x_3+x_1)^2} = \frac{(1-R_0)^2}{(1+R_0)^2} \,, \quad R_0 =\frac{x_1}{x_3}\,,\\
    z_- &=\frac{(x_2-x_1)^2}{(x_2+x_1)^2} = \frac{(1-R_-)^2}{(1+R_-)^2} \,, \quad R_- =\frac{x_1}{x_2} \,. 
\end{split}
\end{align}
Here, for the sake of computational convenience, the cross ratios are defined in three cases, but the final results are written in terms of two cross ratios,  $z_+$ and $z_-$. Additionally, we consider the case where the end of the world brane is orthogonal to the boundary of AdS spacetime, allowing us to ignore its backreaction effects. That is, we set $S_{\mathrm{bdy}} = 0$ and $\theta = 0$ in Fig.~\ref{fig7}.


\paragraph{Disconnected RT configuration.} For the disconnected configuration, there are two disconnected RT surfaces that end at the end of the world brane. One of them starts at the point $x_1$ on the boundary and the other one starts at the point $x_3$ at the boundary. In this case, there are three possible phases of the entanglement wedge cross section (See Fig.~\ref{fig7}). For any phase, the entanglement wedge cross section extends from the point $x_2$ on the boundary of AdS spacetime. Depending on how the other end of the entanglement wedge cross section is located, we have three different phases as follows:

\subparagraph{(i) Phase I.} 
We first consider the phase of the entanglement wedge cross section that anchors at $x_2$ on the boundary of AdS and terminates on the end of the world brane. To compute the first-order correction in $m-1$ of the entanglement wedge cross section, we apply the previously discussed method to this system. 
We consider contributions from two cosmic branes to the entanglement wedge cross section separately and sum them. The result is valid only for the linear order in $m-1$. 

We start with the case $x_2 \rightarrow R_+$ and $x_3 \rightarrow 1$ where $0<R_+<1$. Using the conformal transformation \eqref{conformal transformation}, the points can be rewritten with $\rho$ coordinate as
\begin{equation}
    R_+ \rightarrow \rho_+, \quad 1 \rightarrow \infty.
\end{equation}
The bulk geometry dual to the boundary $(\tau, \rho)$ is the hyperbolic black hole, and the cosmic brane covers the horizon.
The area of the minimal surface connecting $\rho_+$ and the end of the world brane is computed as
\begin{align}
\begin{split}
    \text{Area} &=\log{\left(\frac{\beta}{\pi \delta_\rho}\sinh{\frac{2\pi \rho_+}{\beta}}\right)}\\
    &\sim \log\left(\frac{2 x_2}{\delta_x }\right) +  \frac{1}{2}\left(2+ \frac{1+z_+}{1-z_+} \log z_+  \right)(m-1) + \mathcal{O}((m-1)^2) \,,
\end{split}
\end{align}
where we used $\beta=2\pi m$ and $\rho_+=-\frac{1}{2}\log{z_+}$.
According to the discussion in the previous section, $\delta_\rho$ is translated in terms of $\delta_x$.\footnote{Here, the relevant relation between the two cutoffs near $x_2$ and $\rho_+$ is given by $\delta_\rho=\frac{2x_3 \delta_x}{x_3^2-x_2^2}$ through the similar calculation demonstrated in \eqref{eq_cutoff transf}.}
In the second line, we expand the area around $m =1$ up to linear order.

Additionally, the contribution from the cosmic brane anchored at $x_1$ must be added. Through a global conformal transformation, we start with $x_1 \rightarrow 1$ and $x_2 \rightarrow R_-$ where $0< R_- < 1$.
Similarly, the boundary points can be rewritten in $\rho$ coordinates through the conformal transformation \eqref{conformal transformation}, and the bulk geometry describes the configuration where the cosmic brane envelops the hyperbolic black hole. This configuration is exactly the same as the contribution from the brane anchored at $x_3$ except that the cross ratio $z_+$ is replaced with $z_-$.\footnote{Note that the conformal transformation $x_1\rightarrow 1$ and $x_2\rightarrow R_-$ changes the order of the points. Due to this fact, when we derive the inverse transformation of the cutoff, as in \eqref{eq_cutoff transf}, we need to take into account the change in the order between the point and the cutoff near the point.}

As a result, the entanglement wedge cross section up to the first order of $m-1$ is obtained as 
\begin{equation}\label{eq_EWCS disjoint 1}
         E_{W}^{(m)}(A: B)=\frac{1}{4 G_N} \log{\left(\frac{2 x_2}{\delta_x }\right)}  + \frac{1}{8 G_N} \left( 4 + C_1 (z_+, z_-) \right)  (m-1)  +\mathcal{O}((m-1)^2) \,,
\end{equation}
where
\begin{equation}
  C_1 (z_+, z_-) = \frac{1+z_+}{1-z_+} \log  z_+ +\frac{1+z_-}{1-z_-} \log  z_- \,.
\end{equation}
Note that the leading order term matches the entanglement wedge cross section without cosmic brane for $m=1$ \eqref{eq_appendix phase 1}.

This gravity calculation, along with Fig.~\ref{fig7}, is useful for understanding the corresponding computation in the CFT framework.
The reflected entropy on the CFT side involves a six-point function in its numerator. At leading order, the calculation is related to determining the two-point function since the six-point function does factorize for $m=1$. However, if we consider the $m-1$ correction, it does not factorize. On the gravity side, the origin of this phenomenon can be understood by the influence of the cosmic brane anchored at the boundary points $x_1$ or $x_3$. In the phase I diagram in Fig.~\ref{fig7}, while each cosmic brane affects the entanglement wedge, their influences do not overlap at linear order; instead, the overlap is captured by higher-order corrections. As a result, the reflected entropy, which is related to a six-point function on the CFT side, would be expressed in terms of two four-point functions. This is evident from the expression of $C_1(z_+,z_-)$, where it is written in terms of two cross ratios, but the $z_+$ and $z_-$ function parts are completely separated.

This result provides a holographic framework for analyzing the $\epsilon$-expansion of conformal blocks in six-point correlation functions, for the operator configuration  corresponding to this phase. In particular, the calculation corresponds to the $m-1$ correction to the following six-point correlation function in the CFT\footnote{The relation between the reflected entropy and the correlation function is discussed in \ref{app_A.1}.}
\begin{align}\label{eq_six point function}
    \left\langle  \sigma_{g_B} (-x_3) \sigma_{g_B^{-1}  g_A}(-x_2)  \sigma_{g_A^{-1} }(-x_1)   \sigma_{g_A }(x_1)   \sigma_{g_B g_A^{-1} }(x_2) \sigma_{g_B^{-1}}(x_3)\right\rangle_{\mathrm{CFT}}\,.
\end{align}
This correlator consists of four heavy operators and two light operators in the case where $n=1$ and $m\neq 1$.
The separation of contributions from $z_+$ and $z_-$ in equation \eqref{eq_EWCS disjoint 1} suggests that the $m-1$ correction to the six-point function in the CFT can be effectively captured by the $\epsilon$-expansion of four-point conformal blocks.

\subparagraph{(ii) Phase II.}
The entanglement wedge cross section can also have a phase that anchors at $x_2$ on the boundary of AdS and ends on the brane stretching out from $x_1$. To compute the first-order correction of the entanglement wedge cross section, we consider two contributions: one from the backreaction of the brane at $x_3$ and the other from one at $x_1$.

First, consider the backreaction effect of the brane anchored at $x_3$.
By a global conformal transformation, the points are transformed to 
\begin{equation}
    x_1 \rightarrow R_0, \quad x_2 \rightarrow R_+, \quad x_3 \rightarrow 1 \,,
\end{equation}
where $R_0 < R_+ <1$.
Also, by the conformal transformation \eqref{conformal transformation}, they are rewritten with $\rho$ coordinate as
\begin{equation}
   R_0 \rightarrow \rho_0, \quad R_+ \rightarrow \rho_+, \quad 1 \rightarrow \infty \,.
\end{equation}
Then, the minimal area of the surface connecting $x_2$ and the cosmic brane anchored at $x_1$ in the hyperbolic black hole geometry is computed as\footnote{Here, the relevant relation between the two cutoffs near $x_2$ and $\rho_+$ is given by $\delta_\rho=\frac{2x_3 \delta_x}{x_3^2-x_2^2}$.}
\begin{align}
    \text{Area} &=\log{\left(\frac{\beta}{\pi \delta_\rho} \frac{\cosh{\frac{2\pi \rho_+}{\beta}}-\cosh{\frac{2\pi \rho_0}{\beta}}}{\sinh{\frac{2\pi \rho_0}{\beta}}}\right)}\\
    &\sim \log{\left(\frac{x_2^2 - x_1^2}{x_1 \delta_x}\right)} + \frac{1}{2}\left(2 + C_2 (z_+, z_-) \right) (m-1) + \mathcal{O}((m-1)^2) \,,
\end{align}
where
\begin{equation}
  C_2 (z_+, z_-) = \frac{\left(\sqrt{z_+}+\sqrt{z_-}\right) \left(1+ \sqrt{z_+ z_-} \right) }{ \sqrt{z_-} (1-z_+) (1-z_-)} \left( (1-z_- ) \log z_+  - 2 (1+ z_-) \log \left(\frac{\sqrt{z_+}+\sqrt{z_-}}{1+ \sqrt{z_+ z_-}}\right) \right).
\end{equation}

Next, consider the backreaction effect of the brane anchored at $x_1$. This contribution can be computed from the configuration stretching between the boundary of AdS spacetime and the black hole horizon.
Through a conformal transformation that sets $x_1 \rightarrow 1, x_2 \rightarrow R_-, x_3 \rightarrow R_0$, the area of the minimal surface is obtained as
\begin{align}
    \text{Area} &=\log{\left(\frac{\beta}{\pi \delta_\rho}\right)} \sim \log{\left(\frac{x_2^2 - x_1^2}{x_1 \delta_x }\right)} + (m-1) \,. 
    \label{AreaGadjBHPh2-1st}
\end{align}
As a result, the entanglement wedge cross section up to the first order of $m-1$ is obtained as
\begin{equation}\label{EWphase2}
 E_{W}^{(m)}(A: B)= \frac{1}{4 G_N}\log{\left(\frac{x_2^2 - x_1^2}{x_1 \delta_x}\right)} + \frac{1}{8 G_N} \left( 4 + C_2 (z_+, z_-) \right) (m-1)+ \mathcal{O}((m-1)^2) \,.
\end{equation}
Unfortunately, we do not have results for the Renyi reflected entropy for two intervals, \(A = [x_1, x_2]\) and \(B = [x_2, x_3]\), computed in the BCFT setup. One interesting aspect of our results for the entanglement wedge cross section is that \(C_2(z_+, z_-)\) may be related to an \(\epsilon\)-expansion of five-point functions in the CFT. As can be seen in the phase II diagram in Fig.~\ref{fig7}, the backreaction of the cosmic brane affects both the other cosmic brane and the entanglement wedge cross section, leading to the expectation that the linear order in \(m-1\) is associated with a non-factorized five-point function. The expression for \(C_2(z_+, z_-)\) is given in terms of two cross ratios, \(z_+\) and \(z_-\), and is intricately entangled, unlike in the case of phase I, where the components are separated.

This result suggests that the $m-1$ correction to the entanglement wedge cross section in phase II provides a holographic perspective on the $\epsilon$-expansion of the five-point function in CFT, specifically
\begin{align}\label{eq_five point function}
    \left\langle   \sigma_{g_B} (-x_3) \sigma_{g_A^{-1} }(-x_1)   \sigma_{g_A }(x_1) \sigma_{g_B g_A^{-1} }(x_2) \sigma_{g_B^{-1}}(x_3)  \right\rangle_{\mathrm{CFT}}\,,
\end{align}
which consists of four heavy operators and one light operator in the limit $n=1$ and $m\neq 1$.
In conjunction with the relation between equations \eqref{eq_EWCS disjoint 1} and \eqref{eq_six point function}, this observation indicates that gravity calculations may provide novel insights into the structure of conformal blocks within the holographic framework.

\subparagraph{(iii) Phase III.}
In phase III, the entanglement wedge cross section anchors at $x_2$ on the boundary of AdS and terminates on the brane stretching out from $x_3$. By symmetry, exchanging $x_1$ and $x_3$ in equation (\ref{EWphase2}) provides the expression for phase III. Consequently, the entanglement wedge cross section up to the first order of $m-1$ is obtained as
\begin{equation}
         E_{W}^{(m)}(A: B)= \frac{1}{4 G_N}\log{\left(\frac{x_3^2 - x_2^2}{x_3 \delta_x}\right)}  + \frac{1}{8 G_N} \left( 4 + C_3 (z_+, z_-) \right) (m-1)+ \mathcal{O}((m-1)^2) \,,
\end{equation}
where
\begin{equation}\label{eq_C3}
  C_3 (z_+, z_-) = \frac{\left(\sqrt{z_+}+\sqrt{z_-}\right) \left(1+ \sqrt{z_+ z_-} \right) }{ \sqrt{z_+} (1-z_+) (1-z_-)} \left( (1-z_+) \log z_-  -  2(1+ z_+) \log \left(\frac{\sqrt{z_+}+\sqrt{z_-}}{1+ \sqrt{z_+ z_-}}\right) \right).
\end{equation}
In the case of phase III, only the role of the cross ratios in the expression changes, but the structure remains the same as in phase II.

\paragraph{Connected RT configuration.} For the connected configuration, $x_1$ and $x_3$ are two boundary endpoints of the RT surface. In this case, there is only one phase of the entanglement wedge cross section that starts at the point $x_2$ on the boundary and ends at the RT surface (See Fig.~\ref{fig_GeneralEWCS_Connected}).

Since the entanglement wedge is bounded by the RT surface and the boundary of AdS, the end of the world brane does not influence the entanglement wedge cross section calculation if we disregard the tension of the end of the world brane. Thus, the calculation reduces to that of the entanglement wedge cross section for adjacent two intervals on the boundary in AdS/CFT setup without an end of world brane. The entanglement wedge cross section for adjacent two intervals can be obtained by taking a limit from the entanglement wedge cross section for two disjoint intervals, which is explored in \cite{Jeong:2019xdr}. 

For this reason, let's first consider the entanglement wedge cross section for two disjoint intervals in AdS/CFT. For two disjoint intervals $\bar{A}=[\bar{x_1},\bar{x_2}]$ and $\bar{B}=[\bar{x_3},\bar{x_4}]$ on the boundary of AdS spacetime, there are two possible phases of the entanglement wedge associated with $A\cup B$: (i) the disconnected phase and (ii) the connected phase.\footnote{The terminology is the same as that of AdS/BCFT for general adjacent intervals, but there is a little difference in the meaning of ``connected'' and ``disconnected''.} In the disconnected phase of the entanglement wedge, there are two distinct entanglement wedges of the subregions $\bar{A}$ and $\bar{B}$ respectively. Thus, there is no entanglement wedge cross section as the entanglement wedge is already separated. On the other hand, in the connected phase, a single entanglement wedge is bounded by two RT surfaces connecting the two boundary subregions $\bar{A}$ and $\bar{B}$. In this case, the non-trivial entanglement wedge cross section exists, extending from one RT surface to the other.

For the two disjoint intervals in AdS/CFT setup, the area of the minimal surface connecting two RT surfaces is given by \cite{Jeong:2019xdr}
\begin{equation}\label{eq_CFT entropy 2}
\text{Area}=\log \left[\frac{1+\sqrt{z}}{1-\sqrt{z}}\right]-(m-1) \frac{\sqrt{z} \log z}{1-z}+\mathcal{O}\left((m-1)^2\right) \,,
\end{equation}
where it includes the backreaction from one of the RT surfaces. Here, $z=\frac{\left(\bar{x}_1-\bar{x}_2\right)\left(\bar{x}_3-\bar{x}_4\right)}{\left(\bar{x}_1-\bar{x}_3\right)\left(\bar{x}_2-\bar{x}_4\right)}$ is the cross ratio. This result holds for any $\bar{x}_1, \bar{x}_2, \bar{x}_3,\bar{x}_4$ when the connected phase is dominant. 

To obtain the entanglement wedge cross section for two adjacent intervals, let us consider the neighboring limit of the two intervals $A$ and $B$:
\begin{equation}
(\bar{x}_3-\bar{x}_2) \rightarrow 0 \,.
\end{equation}
Since we are interested in the two adjacent intervals $A=[x_1,x_2]$, $B=[x_2,x_3]$, it is convenient to define the new variables:
\begin{equation}
    \bar{x}_1 \rightarrow x_1,\quad \bar{x}_2 \rightarrow x_2-\delta,\quad \bar{x}_3 \rightarrow x_2+\delta,\quad \bar{x}_4 \rightarrow x_3 \,,
\end{equation}
where we introduced a small $\delta$ in order to concisely define the neighboring limit. Then, the cross ratio is written in terms of the new variables as:
\begin{equation}\label{eq_new cross ratio}
    z=\frac{(x_2-x_1-\delta)(x_3-x_2-\delta)}{(x_2-x_1+\delta)(x_3-x_2+\delta)} \,.
\end{equation}

By substituting \eqref{eq_new cross ratio} into \eqref{eq_CFT entropy 2} and taking the limit $\delta \rightarrow 0$, the entanglement wedge cross section for two adjacent intervals in AdS/CFT setup is obtained by
\begin{align}\label{eq_EWCS connected adj}
    E_{W}^{(m)}(A: B) =\frac{1}{4G_N}\log{\left[\frac{2(x_2-x_1)(x_3-x_2)}{\delta(x_3-x_1)}\right]}+\frac{1}{4G_N}(m-1)+\mathcal{O}\left((m-1)^2\right)\,.
\end{align}
As discussed, for the connected RT configuration of general adjacent intervals in AdS/BCFT setup, the entanglement wedge cross section is equivalent to that of AdS/CFT setup since  the end of the world brane does not contribute. Therefore, \eqref{eq_EWCS connected adj} represents the entanglement wedge cross section involving the $m-1$ correction to for the connected RT configuration in AdS/BCFT. In order to match the zeroth order result in \eqref{eq_A.4}, $\delta$ should equal the UV cutoff $\delta_x$. In the next section, using the same strategy we will reproduce a result from phase II of  adjacent intervals anchored to the boundary of BCFT, as derived in Sec.~\ref{sec_3.2}.

\subsection{Relation to entanglement wedge cross section in AdS$_3$/CFT$_2$}\label{sec_4.2}
In this section, we explore a case where the reflected entropy and the entanglement wedge cross section in AdS$_3$/BCFT$_2$ can be viewed as limiting cases of AdS$_3$/CFT$_2$. We analyze the consistency between our work and \cite{Jeong:2019xdr} by examining this specific case.
It appears that the reflected entropy of AdS$_3$/BCFT$_2$ in phase II of the adjacent intervals in Sec.~\ref{sec_3.2} can be seen as a limiting case of the reflected entropy in the connected phase of AdS$_3$/CFT$_2$.

Let us examine  the connected minimal surfaces of $A\cup B$ where $A=[\bar{x}_1,\bar{x}_2]$ and $B=[\bar{x}_3,\bar{x}_4]$ with $\bar{x}_1<\bar{x}_2<\bar{x}_3<\bar{x}_4$ are two disjoint intervals in CFT$_2$. The area of the entanglement wedge cross section is then determined by equation (3.15) in \cite{Jeong:2019xdr} 
\begin{equation}\label{eq_CFT entropy}
E_{W}^{(m)}(A: B)=\frac{1}{4 G_N} \log \left(\frac{1+\sqrt{z}}{1-\sqrt{z}}\right)-\frac{ 1}{2 G_N} \frac{\sqrt{z} \log z}{1-z} (m-1) +\mathcal{O}\left((m-1)^2\right) \,,
\end{equation}
where the cross ratio is $z=\frac{\left(\bar{x}_1-\bar{x}_2\right)\left(\bar{x}_3-\bar{x}_4\right)}{\left(\bar{x}_1-\bar{x}_3\right)\left(\bar{x}_2-\bar{x}_4\right)}$. This result holds for any $\bar{x}_1, \bar{x}_2, \bar{x}_3,\bar{x}_4$ when the connected phase is dominant. 

\begin{figure}[ ]
    \centering
\includegraphics[width=1\textwidth]{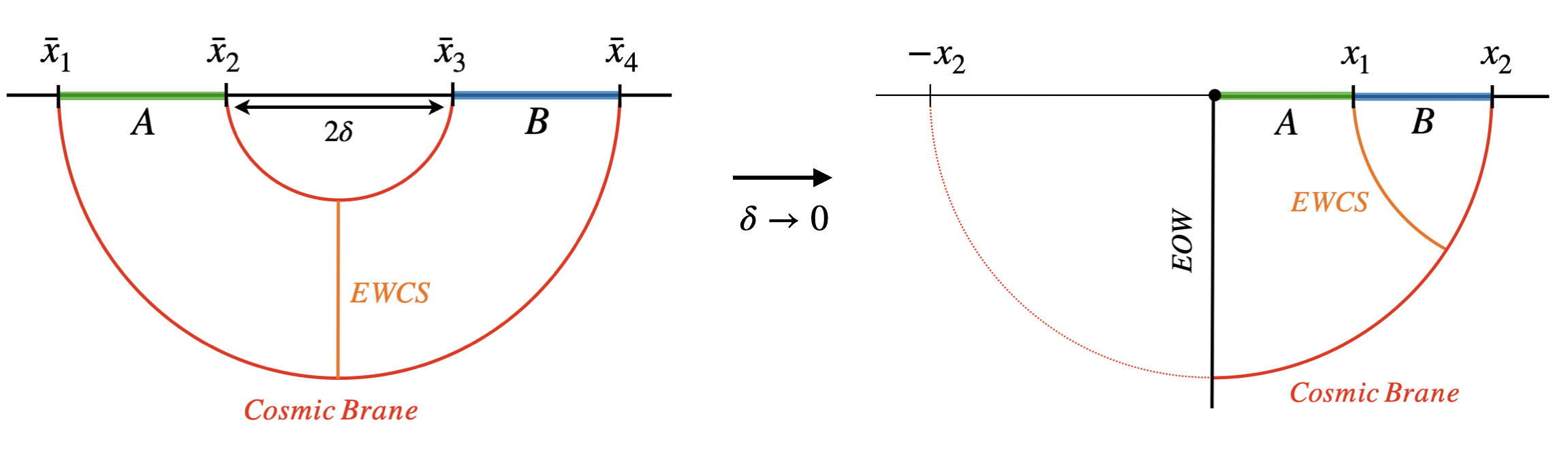}
    \caption{Schematic illustration of the limit from the entanglement wedge cross section in AdS/CFT to the entanglement wedge cross section in phase II in AdS/BCFT. This limit holds only when $|\bar{x}_1|=|\bar{x}_4|$ on the AdS/CFT side and $\theta=0$ on the AdS/BCFT side. The dotted line in the right-hand figure represents a mirror image of the cosmic brane in the geometry dual to the BCFT, motivated by the doubling trick in BCFT.}
    \label{fig_CFTlimit}
\end{figure}

Since the entanglement wedge cross section in AdS$_3$/CFT$_2$ depends on the four points, we need to reduce two degrees of freedom to align with the AdS$_3$/BCFT$_2$ calculation for the adjacent intervals in contact with the boundary of the BCFT, which has two degrees of freedom. To achieve this, we impose the following conditions:
\begin{equation}
\bar{x}_2 \rightarrow \bar{x}_3,\quad |\bar{x}_1|=|\bar{x}_4| \,.
\end{equation}
The first condition, which makes $\bar{x}_2$ coincide with $\bar{x}_3$, reduces one degree of freedom. Motivated by the doubling trick in BCFT (See Fig.~\ref{fig_CFTlimit}), we place $\bar{x}_1$ and $\bar{x}_4$ symmetrically with respect to the origin using the second condition, with $\bar{x}_1<0$ and $\bar{x}_4>0$. In this way, two degrees of freedom remain in total.
To address these conditions conveniently, we define new variables as follows:
\begin{equation}
    \bar{x}_1 \rightarrow -x_2,\quad \bar{x}_2\rightarrow x_1-2\delta,\quad  \bar{x}_3 \rightarrow x_1,\quad \bar{x}_4 \rightarrow x_2 \,,
\end{equation}
where $x_2$ is chosen to satisfy the condition $|\bar{x}_1| = |\bar{x}_4|$. We then move $\bar{x}_2$ close to $\bar{x}_3$, with a small offset of $2\delta$.\footnote{The factor of 2 was introduced for later convenience, as $\delta$ will be related to the cutoff.} Consequently, the cross ratio is redefined as
\begin{equation}
    z=\frac{(x_2-x_1)(x_1+x_2-2\delta)}{(x_1+x_2)(x_2-x_1+2\delta)} \,.
\end{equation}
Taking the $\delta\rightarrow 0$ limit in equation \eqref{eq_CFT entropy}, we obtain
\begin{equation}
\begin{aligned}\label{BCFTPH2}
E_{W}^{(m)}(A: B)=\frac{1}{4 G_N} \log \left(\frac{\left(x_2-x_1\right)\left(x_2+x_1\right)}{ x_2 \, \delta}\right)+\frac{1}{2 G_N}(m-1) + \mathcal{O}((m-1)^2)\,.
\end{aligned}
\end{equation}
This result reproduces the entanglement wedge cross section in phase II of the adjacent intervals anchored to the boundary of BCFT \eqref{EWCSphase2} except for the factor of the first-order correction term. This discrepancy comes from the fact that (\ref{eq_CFT entropy}) includes the contribution from the two cosmic branes, but (\ref{EWCSphase2}) includes the contribution from the one cosmic brane. Here, $\delta$ serves as the UV cutoff $\delta_x$.

\section{Conclusion}\label{sec_conclusion}
In this paper, we calculated the correction term for a Renyi index $m$, one of the two Renyi indices of reflected entropy, in the AdS$_3$/BCFT$_2$ setup. Considering the impact of the boundary of the BCFT, we examined more diverse phases of reflected entropy, and computationally used the doubling trick. On the BCFT side, we calculated the correction term for $m$ in the leading term of the large central charge expansion. In the gravitational system dual to the BCFT, we considered various configurations of the entanglement wedge cross section that could arise from the presence of the end of the world brane. To calculate the backreaction of the entanglement wedge cross section due to the influence of a cosmic brane with tension, we performed a coordinate transformation to convert the Poincare geometry into a hyperbolic black hole geometry, calculated the entanglement wedge cross section, and then obtained the result in terms of the original coordinates. We calculated using different methods in the CFT and AdS, and supported the AdS/CFT correspondence through the agreement of the results from both calculations.

In the case of adjacent intervals that are in contact with the boundary of BCFT, the results match for all orders of $m-1$ due to the characteristics of the configuration. Based on this correspondence, we extended our calculations to a more general configuration of adjacent intervals, allowing for a gap between the boundary at the origin and the subregions of BCFT. For this case, we need to consider an $\epsilon$-expansion of the five-point functions in the CFT, where $\epsilon$ is the rescaled conformal dimension by the central charge, and account for various scenarios. Unfortunately, we were unable to obtain results in the CFT, leaving this for future research. In the corresponding gravitational system, we considered three configurations for general adjacent intervals and calculated the backreaction of the entanglement wedge cross section for each case. Unlike the four-point functions, these are expressed in terms of two cross-ratio values, which could provide insights into how the five-point functions in the CFT are expressed in the $\epsilon$-expansion.

BCFT has been actively studied in the context of entanglement entropy and the black hole information problem. Notably, it plays a central role in reproducing the island prescription within doubly holographic setups and enables modeling black hole evaporation through holographic moving mirror models with time-dependent boundaries \cite{Akal:2020twv, Akal:2021foz, Akal:2022qei, BasakKumar:2022stg}. By introducing boundaries into CFT, one can explore a broader class of physical situations and observe a richer structure of entanglement, including a wider range of configurations and associated phase transitions. Building on our classification of possible phases and the corresponding $m-1$ corrections, our results can be naturally extended to the moving mirror setup to track how the entanglement wedge cross section evolves as the mirror radiates. In this context, the $m-1$ correction to the reflected entropy may serve as a useful probe of non-trivial phase transitions that occur during holographic black hole evaporation, making this an intriguing direction for future research.

Our calculations of the Renyi reflected entropy and the entanglement wedge cross section were performed in Euclidean spacetime. An interesting generalization of our work is to introduce Lorentzian time dependence, such as the dynamics of reflected entropy, pseudo entropy, and timelike entanglement entropy \cite{Kusuki:2019evw, Nakata:2020luh, Doi:2022iyj}. In particular, it is important to investigate in detail the analytic continuation of conformal blocks and surface area in the bulk to the Lorentzian signature.
Another interesting direction is to include quantum corrections \cite{Faulkner:2013ana, Engelhardt:2014gca} to the entanglement wedge cross section, taking into account the backreaction from the cosmic branes.
In this situation, quantum extremal surfaces and island-like contributions may come into play \cite{Dutta:2019gen,Chandrasekaran:2020qtn,Li:2020ceg,Hayden:2021gno,Akers:2022max}.

\acknowledgments
We would like to thank {Hyun-Sik Jeong} for the collaboration at the initial stages of this work. We also thank Jaydeep Kumar Basak and Cynthia Keeler for insightful discussions.
This work was supported by the
Basic Science Research Program through the National Research Foundation of Korea (NRF) funded by the Ministry of Science, ICT \& Future Planning (NRF-2021R1A2C1006791) and the Al-based GIST Research Scientist Project grant funded by the GIST in 2024.
The work was also supported by the National Research Foundation of Korea (NRF) grant funded by the Korea Government (MSIT) (NRF-2023K2A9A1A01095488).
Hospitality at APCTP during the program ``Holography 2024: Correlation and Entanglement in Quantum matter'' is kindly acknowledged.
KYK was also supported by the Creation of the Quantum Information Science R\&D Ecosystem (Grant No.2022M3H3A106307411) through the National Research Foundation of Korea (NRF) funded by the Korean government (Ministry of Science and ICT). 
B. Ahn was supported by Basic Science Research Program through the National Research Foundation of Korea funded by the Ministry of Education (NRF-2020R1A6A3A01095962, NRF-2022R1I1A1A01064342). 
S.-E. Bak is supported by the U.S. Department of Energy under grant number DE-SC0019470 and by the Heising-Simons Foundation “Observational Signatures of Quantum Gravity” collaboration grant 2021-2818.
M.~Nishida was supported by the Basic Science Research Program through the National Research Foundation of Korea (NRF) funded by the Ministry of Education (RS-2023-00245035).  B. Ahn, S.-E. Bak and   M.~Nishida contributed equally to this paper and should be considered
as co-first authors.

\appendix
\section{The $m=1$ calculation for more general adjacent two intervals} \label{app_A}

In this appendix, we summarize the results, without giving details of the calculation, of the reflected entropy and the entanglement wedge cross section with $m=1$ for more general adjacent two intervals in \cite{BasakKumar:2022stg} to make our paper self-contained. For simplicity, we set $S_{\mathrm{bdy}}=0$, which is equivalent to $\theta=0$.

\subsection{Reflected entropy for $m=1$}\label{app_A.1} 
\paragraph{General adjacent intervals.}
We summarize the reflected entropy of two intervals $A=[x_1,x_2]$ and $B=[x_2,x_3]$ at a fixed time slice $t=0$ for $m=1$,
\begin{align}\label{reflected entropy appendix}
    S_R(A: B) = \lim_{m, n \rightarrow 1} \frac{1}{1-n} \log  \frac{\left\langle \sigma_{g_A }(x_1) \sigma_{g_B g_A^{-1} }(x_2) \sigma_{g_B^{-1}}(x_3) \right\rangle_{\mathrm{BCFT}^{\otimes mn}}}{\left(\left\langle \sigma_{g_m}(x_1)  \sigma_{g_m^{-1}}(x_3)\right\rangle_{\mathrm{BCFT}^{\otimes m}}\right)^n}\,.
\end{align}
Using the doubling trick for the general configuration of adjacent intervals, the reflected entropy involves the correlation functions of the associated twist operators in chiral CFT.
This results in various phases depending on the dominant conformal block.
Then, the reflected entropy is determined by taking the minimal values among the results in different phases:
\begin{align}
    S_R(A:B)=\mathrm{min}\left\{S_R^{\text{(phase 1)}},\,S_R^{\text{(phase 2)}},\, S_R^{\text{(phase 3)}},\, S_R^{\text{(phase 4)}}\right\}\,.
\end{align}

\begin{figure}[H]
    \centering
    \includegraphics[width=0.6\textwidth]{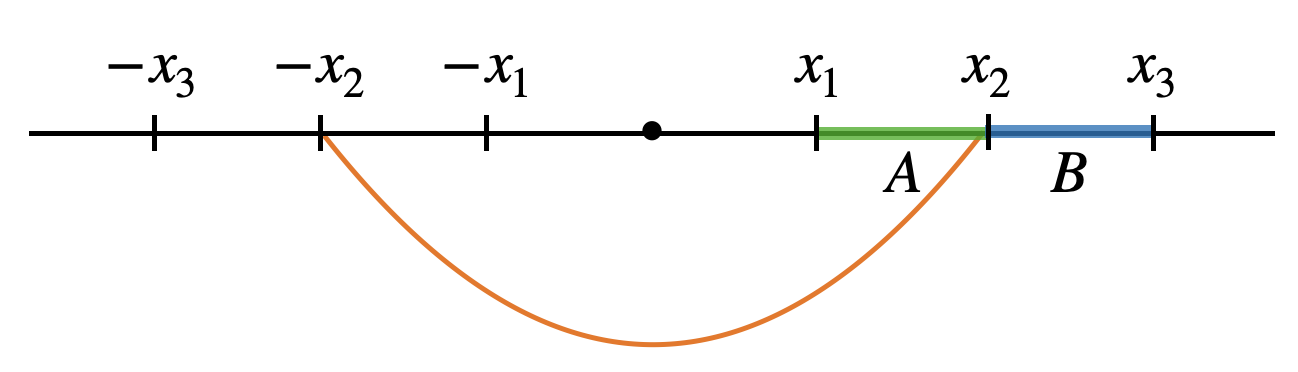}
    \caption{Schematic figure representing the OPE channel for phase I for general adjacent intervals}
    \label{general1}
\end{figure}

\subparagraph{(i) Phase I.}
If the OPE channel takes the form described in Fig.~\ref{general1},
the three-point function in the numerator of equation \eqref{reflected entropy appendix} factorizes as follows:
\begin{align}
    &\left\langle \sigma_{g_A }(x_1) \sigma_{g_B g_A^{-1} }(x_2) \sigma_{g_B^{-1}}(x_3) \right\rangle_{\mathrm{BCFT}^{\otimes mn}} \nn\\
     &= \left\langle  \sigma_{g_B} (-x_3) \sigma_{g_A^{-1} }(-x_1)   \sigma_{g_A }(x_1)  \sigma_{g_B^{-1}}(x_3)   \sigma_{g_B^{-1}  g_A}(-x_2)  \sigma_{g_B g_A^{-1} }(x_2)\right\rangle_{\mathrm{CFT}^{\otimes mn}} \nn\\
    & \sim \left\langle  \sigma_{g_B} (-x_3) \sigma_{g_A^{-1} }(-x_1)   \sigma_{g_A }(x_1)  \sigma_{g_B^{-1}}(x_3) \right\rangle_{\mathrm{CFT}^{\otimes mn}} \left\langle  \sigma_{g_B^{-1}  g_A}(-x_2)  \sigma_{g_B g_A^{-1} }(x_2)\right\rangle_{\mathrm{CFT}^{\otimes mn}}.
\end{align}
The correlation function in the denominator of the equation \eqref{reflected entropy appendix} can be similarly doubled and factorized. Then, the reflected entropy can be expressed in terms of the twist operators in the configuration we consider,
\begin{align}
    S_R(A: B) = \lim_{m, n \rightarrow 1} \frac{1}{1-n} \log  \left\langle  \sigma_{g_B^{-1}  g_A}(-x_2)  \sigma_{g_B g_A^{-1} }(x_2)\right\rangle_{\mathrm{CFT}^{\otimes mn}}.
\end{align}
Consequently, the reflected entropy between two adjacent subregions $A$ and $B$ is given by
\begin{align}
    S_R(A: B)=\frac{c}{3} \log \left(\frac{2 x_2}{\delta_x}\right)+2 S_{\mathrm{bdy}} \,,
\end{align}
where $\delta_x$ is the UV cutoff and $S_{\mathrm{bdy}} $ is the boundary entropy (or logarithm of the g-function).

\begin{figure}[H]
    \centering
    \includegraphics[width=1\textwidth]{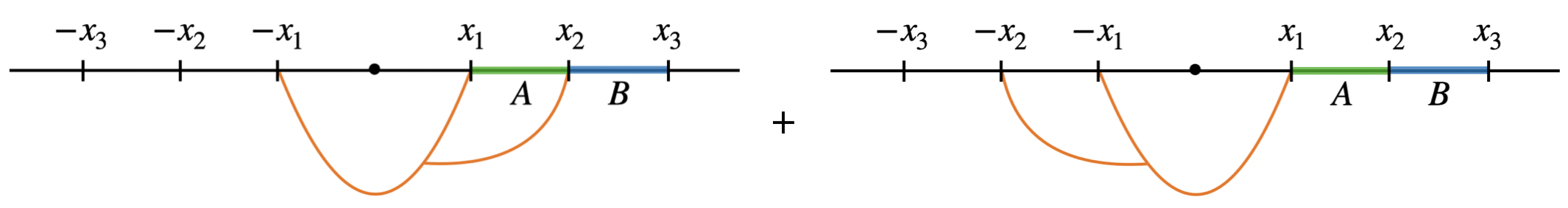}
    \caption{Schematic figure representing the OPE channel for phase II for the general adjacent intervals}
    \label{general2}
\end{figure}

\subparagraph{(ii) Phase II.}
The OPE channel can take the following form:
\begin{align}
    S_R(A: B) &= \lim_{m, n \rightarrow 1} \frac{1}{1-n} \log  \frac{\left\langle   \sigma_{g_A^{-1} }(-x_1)   \sigma_{g_A }(x_1) \sigma_{g_B g_A^{-1} }(x_2)  \right\rangle \left\langle  \sigma_{g_B} (-x_3) \sigma_{g_B^{-1}}(x_3)  \right\rangle_{\mathrm{CFT}^{\otimes mn}}}{\left(\left\langle  \sigma_{g_m^{-1}}(-x_1) \sigma_{g_m}(x_1)   \right\rangle \left\langle \sigma_{g_m}(-x_3)   \sigma_{g_m^{-1}}(x_3) \right\rangle_{\mathrm{CFT}^{\otimes m}}\right)^n}\, \nonumber\\
    &= \lim_{m, n \rightarrow 1} \frac{1}{1-n} \log  \frac{\left\langle   \sigma_{g_A^{-1} }(-x_1)   \sigma_{g_A }(x_1) \sigma_{g_B g_A^{-1} }(x_2)  \right\rangle_{\mathrm{CFT}^{\otimes mn}}}{\left(\left\langle  \sigma_{g_m^{-1}}(-x_1) \sigma_{g_m}(x_1)   \right\rangle_{\mathrm{CFT}^{\otimes m}}\right)^n}.
\end{align}
In this phase, there is another contribution from $\langle   \sigma_{g_B^{-1} g_A }(-x_2) \sigma_{g_A^{-1} }(-x_1)   \sigma_{g_A }(x_1)   \rangle$ in the chiral CFT, as depicted in Fig.~\ref{general2}. As a result, the reflected entropy between two adjacent subregions $A$ and $B$ is given by
\begin{align}
    S_R(A:B)=\frac{c}{3}\log{\left(\frac{(x_1-x_2)(x_1+x_2)}{x_1\,\delta_x}\right)} \,.
\end{align}

\begin{figure}[H]
    \centering
    \includegraphics[width=1\textwidth]{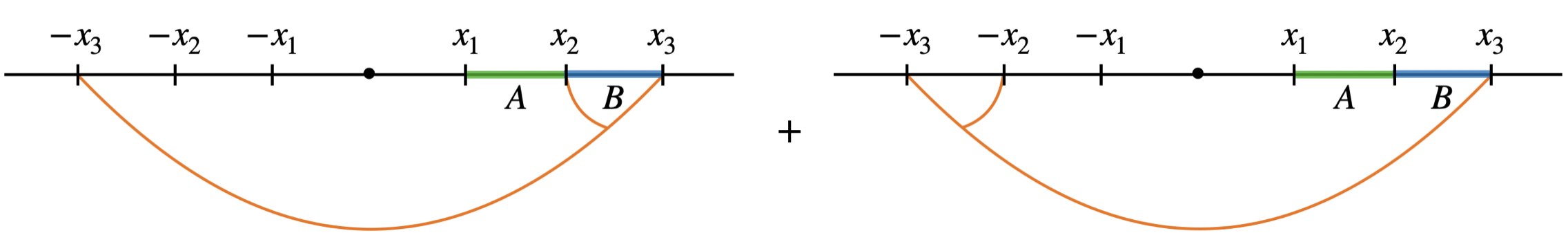}
    \caption{Schematic figure representing the OPE channel for phase III for the general adjacent intervals}
    \label{general3}
\end{figure}

\subparagraph{(iii) Phase III.}
If the OPE channel takes the following form
\begin{align}
    S_R(A: B) &= \lim_{m, n \rightarrow 1} \frac{1}{1-n} \log  \frac{\left\langle  \sigma_{g_B} (-x_3) \sigma_{g_B g_A^{-1} }(x_2)  \sigma_{g_B^{-1}}(x_3)  \right\rangle_{\mathrm{CFT}^{\otimes mn}}}{\left(\left\langle \sigma_{g_m}(-x_3)   \sigma_{g_m^{-1}}(x_3) \right\rangle_{\mathrm{CFT}^{\otimes m}}\right)^n}\,,
\end{align}
with another contribution $\langle  \sigma_{g_B} (-x_3) \sigma_{g_B g_A^{-1} }(x_2)  \sigma_{g_B^{-1}}(x_3) \rangle$ as depicted in Fig.~\ref{general3}, then the reflected entropy between two adjacent subregions $A$ and $B$ is obtained as
\begin{align}
    S_R(A:B)=\frac{c}{3}\log{\left(\frac{(x_3-x_2)(x_3+x_2)}{x_3\,\delta_x}\right)} \,.
\end{align}

\begin{figure}[H]
    \centering
    \includegraphics[width=1.0\textwidth]{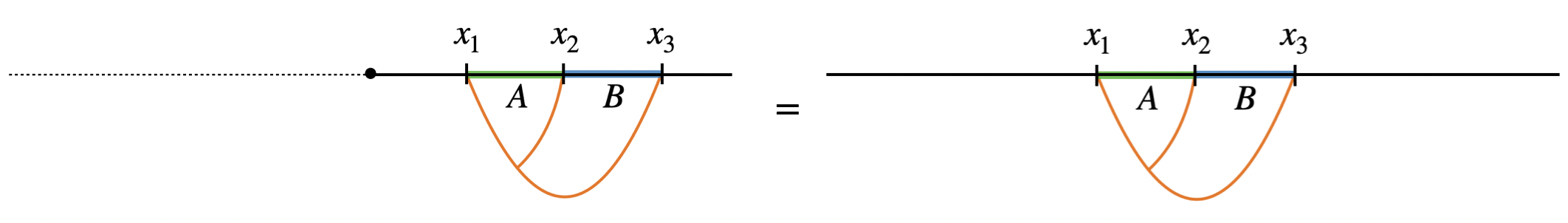}
    \caption{Schematic figure representing the OPE channel for phase IV for the general adjacent intervals. In this configuration, there is no contribution from the boundary at the origin, so the three-point function of BCFT is the same with that of CFT.}
    \label{general4}
\end{figure}
\subparagraph{(iv) Phase IV.} If the OPE channel doesn't involve the contribution from the boundary, the reflected entropy in BCFT is the same as the result obtained from the three-point function in CFT (See Fig.~\ref{general4}):
\begin{align}
    S_R(A: B) &= \lim_{m, n \rightarrow 1} \frac{1}{1-n} \log  \frac{\left\langle \sigma_{g_A }(x_1) \sigma_{g_B g_A^{-1} }(x_2) \sigma_{g_B^{-1}}(x_3) \right\rangle_{\mathrm{BCFT}^{\otimes mn}}}{\left(\left\langle \sigma_{g_m}(x_1)  \sigma_{g_m^{-1}}(x_3)\right\rangle_{\mathrm{BCFT}^{\otimes m}}\right)^n}\\
    &= \lim_{m, n \rightarrow 1} \frac{1}{1-n} \log  \frac{\left\langle \sigma_{g_A }(x_1) \sigma_{g_B g_A^{-1} }(x_2) \sigma_{g_B^{-1}}(x_3) \right\rangle_{\mathrm{CFT}^{\otimes mn}}}{\left(\left\langle \sigma_{g_m}(x_1)  \sigma_{g_m^{-1}}(x_3)\right\rangle_{\mathrm{CFT}^{\otimes m}}\right)^n}\,.
\end{align}
We then obtain the reflected entropy as
\begin{align}\label{eq_phase 4 appendix}
    S_R(A:B) =\frac{c}{3} \log{\left(\frac{2(x_3-x_2)(x_2-x_1)}{(x_3-x_1) \delta_x}\right)}  \,.
\end{align}
\subsection{Entanglement wedge cross section for $m=1$}\label{app_A.2}
We summarize the entanglement wedge cross section of two intervals $A=[x_1,x_2]$ and $B=[x_2,x_3]$ at a fixed time slice $t=0$ for $m=1$.
For $m=1$, the entanglement wedge cross section is obtained by calculating the geodesic length of each case in the geometry without backreaction of the cosmic brane. In principle, this is the same kind of calculation as in Sec.~\ref{sec_3.1.2}.
As we discussed in Sec.~\ref{sec 4.1}, there are disconnected and connected RT configurations for the subregion $A\cup B$.

\paragraph{Disconnected RT configuration.} For the disconnected RT configuration, there are three possible phases of the entanglement wedge cross section (See Fig.~\ref{fig7}).
\subparagraph{(i) Phase I.} If the end of the entanglement wedge cross section is located on the end of the world brane, the geodesic length is given by the same calculation from \eqref{eq_zeroth EW simple 1} except that $x_2$ is the point where the entanglement wedge cross section anchors 
\begin{align}\label{eq_appendix phase 1}
    E_W(A: B)=\frac{1}{4 G_N}\log{\left(\frac{2x_2}{\delta_x}\right)} \,,
\end{align}
where $\delta_x$ is the UV cutoff.
\subparagraph{(ii) Phase II.} If the end of the entanglement wedge cross section is located on the RT minimal surface associated with the boundary point $x_1$, the entanglement wedge cross section is given by\footnote{We flipped $x_1\leftrightarrow x_2$ in \eqref{eq_zeroth EW simple 2} while keeping the variable of the logarithmic function positive to obtain \eqref{eq_zeroth EW general 2}.}
\begin{align}\label{eq_zeroth EW general 2}
    E_W(A: B) =\frac{1}{4G_N} \log{\left(\frac{(x_2-x_1)(x_1+x_2)}{x_1\delta_x}\right)}  \,.
\end{align}

\subparagraph{(iii) Phase III.} If the end of the entanglement wedge cross section is located on the RT minimal surface associated with the boundary point $x_3$, the entanglement wedge cross section is given by\footnote{We replaced $x_1$ in \eqref{eq_zeroth EW simple 2} with $x_2$ and replaced $x_2$ in \eqref{eq_zeroth EW simple 2} with $x_3$ to obtain \eqref{eq_zeroth EW general 3}.} 
\begin{align}\label{eq_zeroth EW general 3}
    E_W(A: B) =\frac{1}{4G_N} \log{\left(\frac{(x_3-x_2)(x_3+x_2)}{x_3 \delta_x}\right)}  \,.
\end{align}

\paragraph{Connected RT configuration.} For the connected configuration, $x_1$ and $x_3$ are two boundary end points of the RT surface. In this case, there is only one phase of the entanglement wedge cross section that extends from the point $x_2$ on the boundary and ends at the RT surface. (See Fig.~\ref{fig_GeneralEWCS_Connected}). The entanglement wedge cross section is given by
\begin{align}\label{eq_A.4}
    E_W(A: B) =\frac{1}{4G_N} \log{\left(\frac{2(x_2-x_1)(x_3-x_2)}{(x_3-x_1) \delta_x}\right)}  \,,
\end{align}
where $\delta_x$ is the UV cutoff. The entanglement wedge cross section in the connected RT configuration is dual to the reflected entropy in phase IV \eqref{eq_phase 4 appendix}.

\bibliographystyle{JHEP}

\end{document}